\documentclass[twocolumn,preprint]{aastex62}
\usepackage[T1]{fontenc}
\usepackage[english]{babel}
\usepackage{natbib}
\usepackage{ulem,xcolor}
\usepackage{amsmath, graphicx}
\usepackage{soul}
\def\rhessi{{\textit{RHESSI}}}
\def\IRIS{{\textit{IRIS}}}
\def\goes{{\textit{GOES}}}

\def\kw{{Konus-\textit{Wind}}}
\def\sdo{{\textit{SDO}}}
\def\hinode{{\textit{Hinode}}}
\def\mw{{microwave}}

\def\gs{{gyrosynchrotron}}

\begin{document}

\title{Energy budget of plasma motions, heating, and electron acceleration in a three-loop solar flare}

	\author[0000-0001-5557-2100]{Gregory D. Fleishman}
	\affil{Center For Solar-Terrestrial Research, New Jersey Institute of Technology, Newark, NJ 07102, USA
	\\
		Ioffe Institute, Polytekhnicheskaya, 26, St. Petersburg, 194021, Russia}

\author[0000-0002-7791-3241]{Lucia Kleint}
	\affil{Leibniz-Institut f\"ur Sonnenphysik (KIS), Sch\"oneckstrasse 6, D-79104 Freiburg, Germany \\
	University of Geneva, CUI, 7 route de Drize, 1227 Carouge, Switzerland}

\author[0000-0001-7856-084X]{Galina G. Motorina}
	\affil{Astronomical Institute of the Czech Academy of Sciences, 251 65 Ond\v{r}ejov, Czech Republic\\
Central  Astronomical Observatory at Pulkovo of Russian Academy of Sciences, St. Petersburg, 196140, Russia}

\author[0000-0003-2846-2453]{Gelu M. Nita}
	\affil{Center For Solar-Terrestrial Research, New Jersey Institute of Technology, Newark, NJ 07102, USA}

	\author[0000-0002-8078-0902]{Eduard P. Kontar}
	\affil{School of Physics \& Astronomy, University of Glasgow, G12 8QQ,
Glasgow, United Kingdom}

\begin{abstract}
Non-potential magnetic energy promptly released in solar flares is converted to other forms of energy.
This may include nonthermal energy of flare-accelerated particles,
thermal energy of heated flaring plasma, and kinetic energy of eruptions, jets, up/down flows, and stochastic (turbulent) plasma motions.
The processes or parameters governing partitioning  of the released energy between these components is an open question. How these components are distributed between distinct flaring loops and what controls these spatial distributions is also unclear.
Here, based on multi-wavelength data and 3D modeling, we quantify the energy partitioning and spatial distribution in the  well observed SOL2014-02-16T064620 solar flare of class C1.5. Nonthermal emissions of this flare displayed a simple impulsive single-spike light curves lasting about 20\,s. In contrast, the thermal emission demonstrated at least three distinct heating episodes, only one of which was associated with the nonthermal component. The flare was accompanied by up and down flows and substantial turbulent velocities.
The results of our analysis suggest that (i) the flare occurs in a multi-loop system that included at least three distinct flux tubes;
(ii) the {released} magnetic energy is divided unevenly between the thermal and nonthermal components in these loops; (iii) only one of these three flaring loops contains an energetically important amount of nonthermal electrons, while two other loops remain thermal;
(iv) the amounts of direct plasma heating and that due to nonthermal electron loss are comparable; (v)  the 
kinetic energy in the flare footpoints constitute only a minor fraction compared with the thermal and nonthermal energies.
\end{abstract}


\section{Introduction}
\label{S_Intro}

Solar flares are explosive phenomena that cover a range of heights in the solar atmosphere. Flares are observed as transient brightenings throughout the electromagnetic spectrum. Some of the brightenings can show apparent motions, while others can appear immobile. Solar flares can be associated with large-scale coronal mass ejections (CMEs), collimated jets, and/or solar energetic particles (SEPs) detected at the heliosphere. All these observables are associated with underlying magnetic, nonthermal, thermal, kinetic, and potential energies, whose transformation chains, partitions, spatial distributions, and temporal evolution are of primary importance for understanding the solar flare phenomenon.

\citet{2012ApJ...759...71E} analyzed the energetics of 38 
eruptive solar flares and, in particular, concluded that
(i) the energy of flare-accelerated charged particles 
exceeds 
the bolometric energy radiated across all wavelengths and, thus, sufficient to supply it; (ii) 
the electrons and ions accelerated in the flare 
gain comparable amounts of energy; and (iii) the free magnetic energy available in the given active region is sufficient to drive the particle acceleration, plasma heating, and power the CME. Although these are important findings that confirm our overall understanding of solar flare energy budget, the flare energy estimates (i) have large uncertainties; (ii) likely contain bias related to selection of a set of rather strong flares; (iii) do not provide unambiguous temporal relationships between various components; and (iv) do not include the kinetic energy of turbulence or bulk motions of the thermal plasma in the flaring loops. Although the turbulent energy can be evaluated using the spectral broadening of relevant spectral lines \citep[e.g.,][and references therein]{2017PhRvL.118o5101K, 2018ApJ...854..122W}, removing or at least minimizing other limitations requires careful case studies  of dissimilar eruptive and confined events. This is particularly important, because the ion acceleration can only be quantified in rather powerful events, where gamma-ray line emission can be detected. The electron acceleration can be detected in a much broader range of flare sizes, but the characterization of the nonthermal electron energy content is often limited because of uncertainty in quantification of the poorly constrained low-energy cut-off in the nonthermal electron spectrum. A more reliable way of the low-energy cut-off finding has been proposed \citep{2019ApJ...871..225K} via a warm-target model, which, however, relies on a number of model assumptions.

Recently, \citet{2018ApJ...856..111L} described a class of early impulsive ``cold'' flares, where the direct plasma heating is weak or nonexistent, while most of the plasma thermal manifestations is driven by impact of nonthermal electrons accelerated in the course of the flare. Given that the thermal emissions from the cold flares is somewhat low, the nonthermal X-ray emission dominates the spectrum down to low energies, which permits much lower low-energy cut-off values to be derived, compared with a typical flare, even from the cold-target fit. The thermal component can also be quantified much more conclusively in the cold flares compared with the normal flare. Indeed, a reasonably low-temperature ($\sim10$\,MK) component of the thermal plasma in a cold flare can be spatially resolved with extreme ultraviolet (EUV) data and, thus, studied in much greater detail than the hotter flaring plasma
visible in X-rays.

A few case studies of the cold flares reported that the cold flares originate due to interaction between two magnetic flux tubes. For example, \citet{2016ApJ...822...71F} analyzed a cold flare with a delayed heating and found that the flare occurred due to interaction between two magnetic flux tubes with strikingly different sizes---one small and one big loop. However, in spite of this remarkable difference in the loop sizes, the nonthermal accelerated electrons divided between these two loops in comparable amounts.
Intriguingly, the thermal response of these two loops on the comparable nonthermal electron impact was different, likely, because of the dissimilar geometry of the loops. In contrast, \cite{2020ApJ...890...75M} demonstrated that another nonthermal-dominated cold flare occurred due to interaction between two flux tubes with comparable sizes. In this case, the nonthermal electrons were also divided roughly equally between these two loops, which might indicate that the nonthermal electron partition between two interacting loops is controlled by local properties of this interaction, rather than the loop sizes. However, the thermal responses of those two loops were again different, likely, because of differences in the thermal plasma properties in the  loops just before the flare.

Here we investigate the SOL2014-02-16T064600 flare that shares a number of properties with the cold flares, although it does not pass the formal criterion for the early-impulsive cold flares proposed by \citet{2018ApJ...856..111L}.
This event is a rare (perhaps, the only) case when a rather simple flare with a `single-spike' impulsive phase was observed with \IRIS\,
so the kinetic energy of turbulent and bulk plasma motions at the flaring loop footpoints could be quantified based on the spectral line analysis. A study of energy distribution, partitioning, and evolution is facilitated by a unique combination of the complementary data sources. However, some of the essential data sets have noticeable temporal gaps, which we attempted to fill with the data-constrained 3D modeling \citep[cf.][]{2020ApJ...890...75M}.

We found that three distinct flux tubes are involved in the SOL2014-02-16T064600 flare. 
However, unlike the cold flares studied, two of these three flaring flux tubes do not show any evidence of a significant nonthermal component.
The nonthermal electron population is only detectable in the largest and hottest loop; this loop showed thermal-to-nonthermal behavior consistent with the Neupert effect \citep{1968ApJ...153L..59N}. The loops, however, demonstrate a noticeable pre-heating phase, with no signature of any nonthermal electron component. The estimated input of the nonthermal electron energy is insufficient to fuel the overall thermal response in this flare, thus favoring an additional mechanism of plasma heating. 
The kinetic energies of the bulk and turbulent motions in the flare footpoints estimated from the shift and width of the spectral lines appear to be much smaller than either thermal or nonthermal energy of the flare.



\section{Observations} \label{S_Observations}

The solar flare SOL2014-02-16T064600, GOES class C1.5,
occurred at $\sim$06:46~UT in AR 11974 with $\beta\gamma$-configuration located at W56S12 (cosine of heliocentric angle $\mu=0.55$). 
The flare displayed a short impulsive profile with a single peak at $\sim$06:44:38\,UT in hard X-ray (HXR) above 20\,keV and in \mw s that lasted about 20\,s.  \\\\

\begin{figure}\centering
\includegraphics[width=0.98\columnwidth]{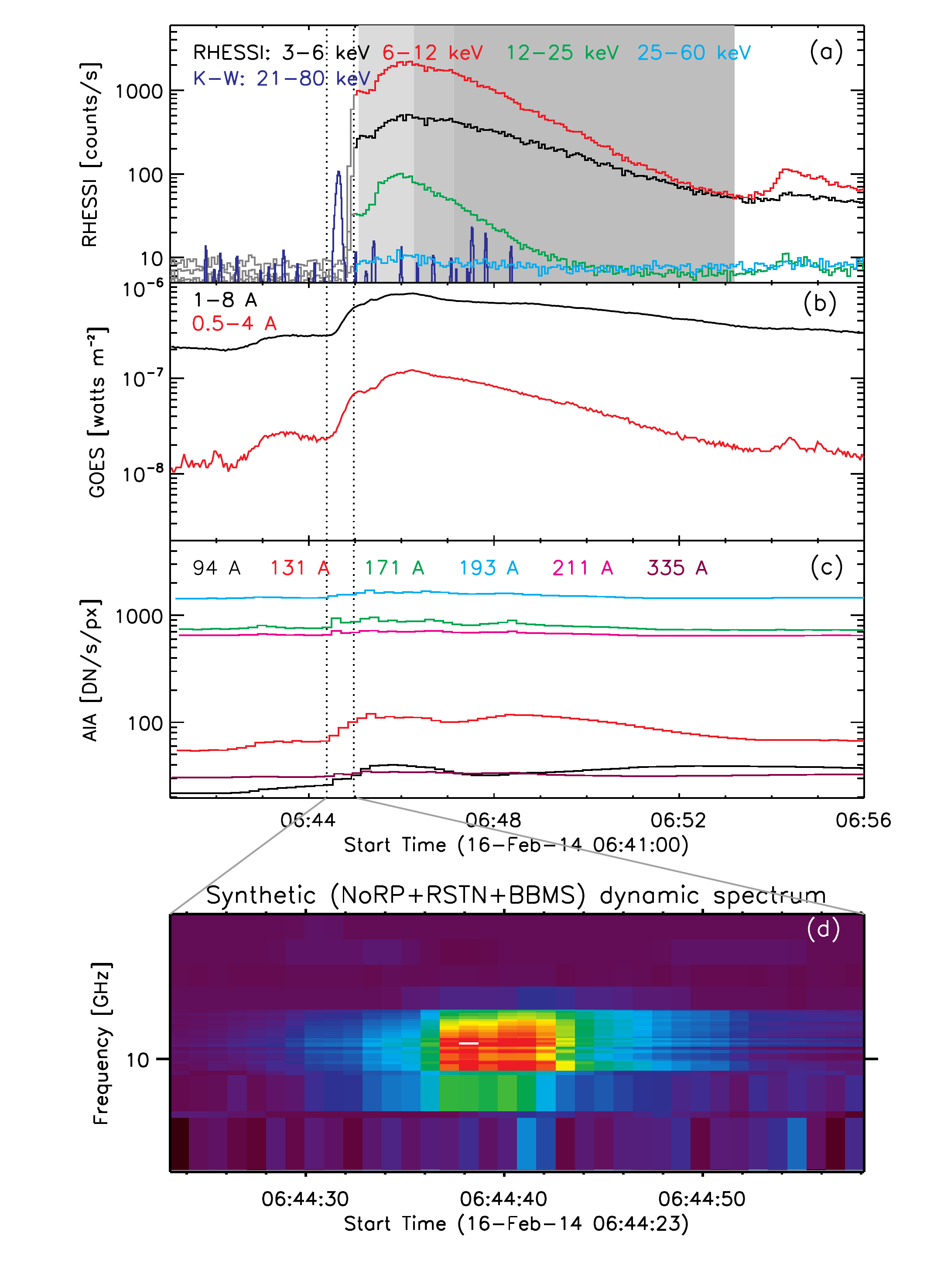}
\caption{Overview of the February 16, 2014 flare. (a) \rhessi\ and \kw\ light curves; \rhessi\ light curves during the orbital night are shown in light gray; (b) \goes\ light curves; 
(c) the AIA light curves obtained from the selected ROI (see Fig.~\ref{Fig_AIA_maps}); (d) NoRP$+$RSTN$+$BBMS dynamic spectrum  of the impulsive flare phase (the absolute peak of the radio flux density is 50\,sfu). Vertical dotted lines indicate the impulsive phase shown in the bottom panel in the \mw\ emission. The light gray area starting on 06:45:06\,UT after the exit from the \rhessi\ night shows the 8s fitted time intervals, the dark gray---the 12s fitted time intervals, the darkest gray--- the 20s fitted time intervals of the \rhessi\ observations. 
\label{Fig_lightcurves}
}
\end{figure}

\subsection{Overview of the instruments used in the analysis}

The flare is observed with a unique combination of space- and ground-based instruments throughout the entire electromagnetic spectrum from radio waves to HXRs, see Figure~\ref{Fig_lightcurves}. Nevertheless, there are substantial gaps in these data, which we attempt to fill with 3D modeling.

In the HXR domain the event is observed by a few instruments: \textit{Reuven Ramaty High Energy Solar Spectroscopic Imager} \citep[\rhessi ,][]{2002SoPh..210....3L}, \kw\; \citep{1995SSRv...71..265A, 2014Ge&Ae..54..943P}, and X-Ray Telescope \citep[XRT,][]{2007SoPh..243...63G} on board the \textit{Hinode} mission  \citep{2007SoPh..243....3K}.
\rhessi\ missed the impulsive phase due to the orbital night, and recorded only the thermal response phase (see Figure \ref{Fig_lightcurves}(a)). \kw\ was not in triggered mode,
so only low-resolution G1 (21-80~keV) light curve is available
(dark blue line, Figure \ref{Fig_lightcurves}a).

\begin{figure}\centering
\includegraphics[width=0.98\columnwidth]{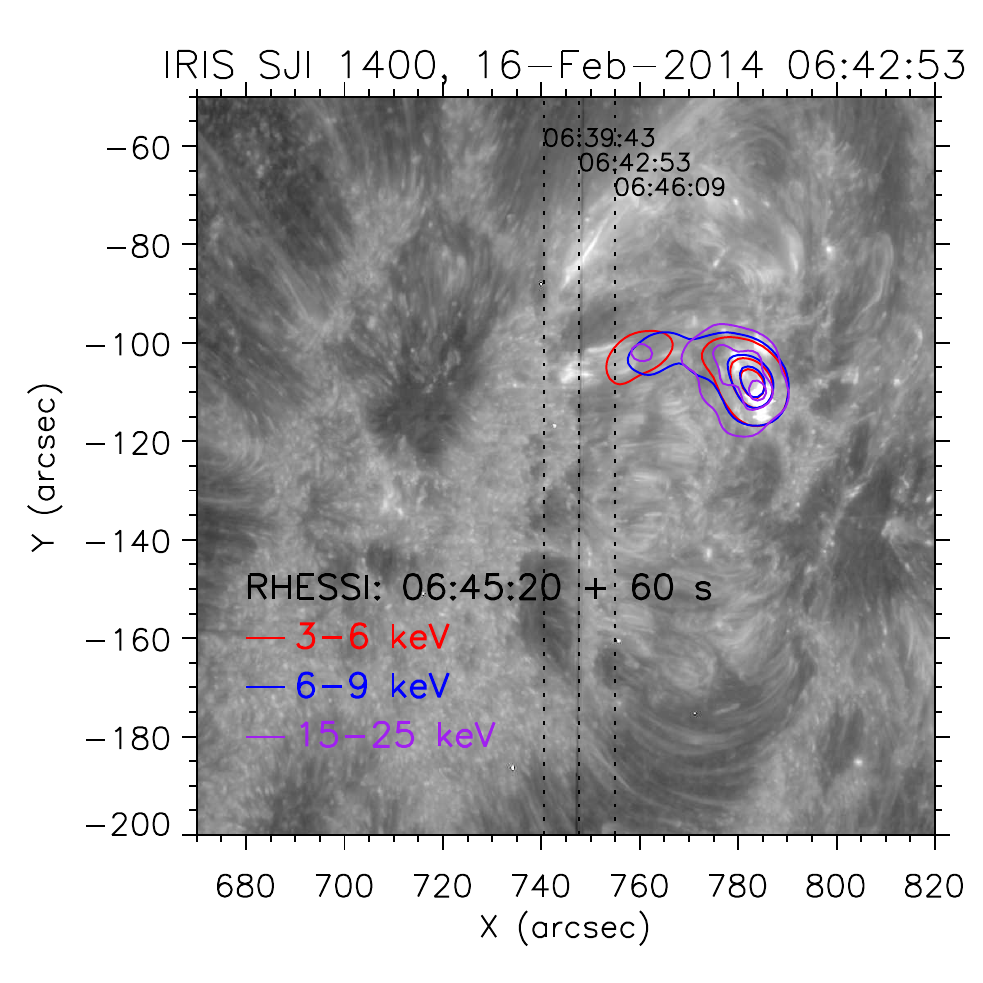}
\caption{The background shows an \IRIS\ SJ 1400 \AA\ image taken at 06:42:54 UT, just as small brightenings were starting in the region. The \rhessi\ source, reconstructed at three different energy bins during a time interval of 60 seconds, overlays some of the brightenings. 
\rhessi\ contours are rotated  0.2$^\circ$ clockwise about disk center as explained in Section\,\ref{S_xray}.
The vertical dotted lines indicate three different \IRIS\ spectrograph slit positions with their times labeled, to show the motion of the spectrograph slit across the solar surface.
\label{irisoverview}}
\end{figure}

\IRIS\ \citep{2014iris} carried out a 400-step raster scan on February 16, 2014 from 06:16--07:19 UT.
Each raster step took 9.5\,seconds, consisting of 8\,s exposure time plus overhead and encompassed an area of $0\farcs33 \times 174\arcsec$. This gives a total FOV of $141\arcsec \times 174\arcsec$, with the caveat that the different slit positions are not obtained simultaneously. Fortunately, the \IRIS\ slit was just above the region of activity at the time of the flare. For each slit position, \IRIS\ recorded near UV (NUV) and far UV (FUV) spectra in several spectral lines that are described below. Figure~\ref{irisoverview} shows a context image and the \IRIS\ slit positions at different time steps.

\begin{figure*}\centering
\includegraphics[width=0.85\textwidth]{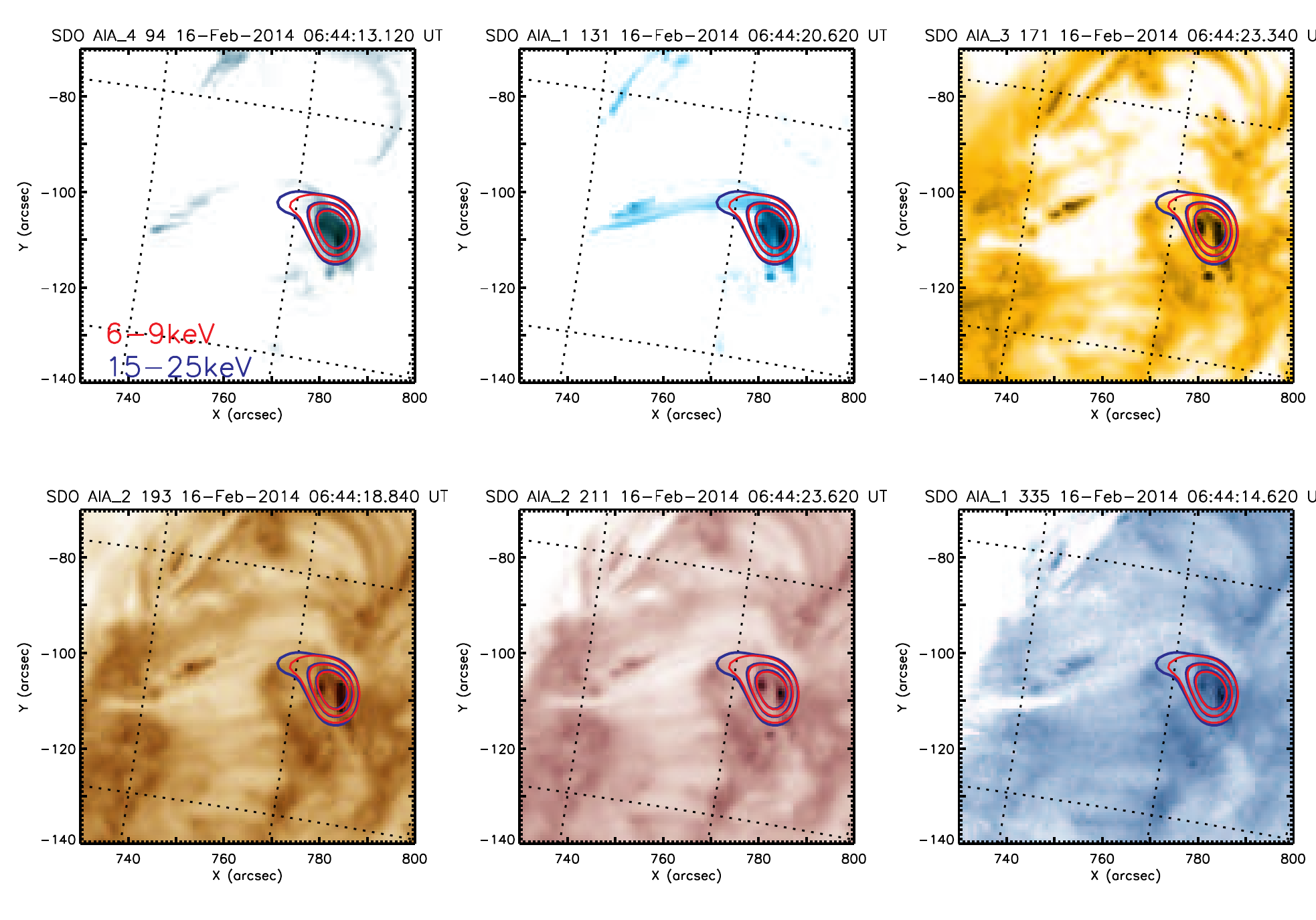}
\caption{AIA maps of the February 16, 2014 flare 
with overlaid \rhessi\ CLEAN image
(Clean Beam Width Factor is CBWF$=1.8$) 30, 50, and 70\% contours for 6-9 keV (red lines) and 15-25 keV (blue lines) for time interval 06:45:20 - 06:46:20~UT. For co-alignment the \rhessi\ roll angle rhessi\_roll\_angle=-0.2  has been applied to rhessi\_roll\_center=(0, 0).
\label{Fig_AIA_maps}
}
\end{figure*}

Soft X-ray are available from the \textit{Geostationary Operational Environmental Satellite} \citep[\goes ,][]{2005SoPh..227..231W} (Figure \ref{Fig_lightcurves}(b)), while optical and EUV data are available from \textit{Solar Dynamics Observatory}/Helioseismic and Magnetic Imager \citep[\sdo/HMI,][]{2012SoPh..275..207S} and  \sdo/Atmospheric Imaging Assembly \citep[\sdo/AIA,][]{2012SoPh..275...17L} respectively. The EUV light curves intergated over the region of interest (ROI, see Figure \ref{Fig_AIA_maps}) are shown in Figure \ref{Fig_lightcurves}(c).

In the \mw\ domain the flare occurred in the time range covered by Siberian Solar Radio Telescope \citep[SSRT,][]{2003SoPh..216..239G} and Nobeyama instruments. 
Nobeyama Radioheliograph \citep[NoRH,][]{1994IEEEP..82..705N} finished observations a few minutes before the event.  SSRT was in a transition mode to a new instrument: 
the data were taken but no calibration has been available at the time of the flare to produce images. The Solar Radio Spectropolarimeters \citep[SRS,][]{Muratov2011}  2-24 GHz data were lost because of disk failure (A.T. Altyntsev; private communication). The available \mw\ data set includes Nobeyama Radio Polarimeter \citep[NoRP,][]{1979PRIAN..26..129T} data at a few frequencies, Radio Solar Telescope Network \citep[RSTN,][]{1981BAAS...13Q.553G}, and the Badary Broadband Microwave Spectropolarimeters \citep[BBMS,][]{2015SoPh..290..287Z} data at 4-8\, GHz,  the combined dynamic spectrum is shown in Figure \ref{Fig_lightcurves}(d).
\newpage

\subsection{X-ray data}
\label{S_xray}

Hard X-ray observations of the flare impulsive phase are only available in the \kw\ wide G1 channel covering 21--80\,keV range. The light curve with the time cadence of 2.994\,s recorded in the waiting mode, which we combine with the \mw\ light curves to evaluate the nonthermal electron escape time from the radio source.

\rhessi\ HXR observations with high temporal (2 s)
and energy (1 keV) resolution provide information on electrons in the range from $\sim$3 keV up to $\sim$30-50 keV, as well as maps of X-ray sources with a spatial resolution of $\sim$7" are only available in the decay phase of the flare.
\rhessi\ data available after the terminator $\approx 06:45$~UT
are used to produce images in  the flare decay phase at a few spectral intervals between 3 and 25\,keV using CLEAN algorithm \citep{2002SoPh..210...61H} with Clean Beam Width Factor CBWF$=1.8$; see Figure\,\ref{Fig_AIA_maps}.

A loop-like structure connecting foot-points is observed during the impulsive peak with the \hinode/XRT using various filters.
The \rhessi\ images were rotated by the roll angle rhessi\_roll\_angle=-0.2  using rhessi\_roll\_center=(0, 0) to co-align with these \hinode/XRT images, which are properly co-aligned with the \sdo/AIA data.


%
%
%

\begin{figure*}\centering
\includegraphics[width=0.75\textwidth]{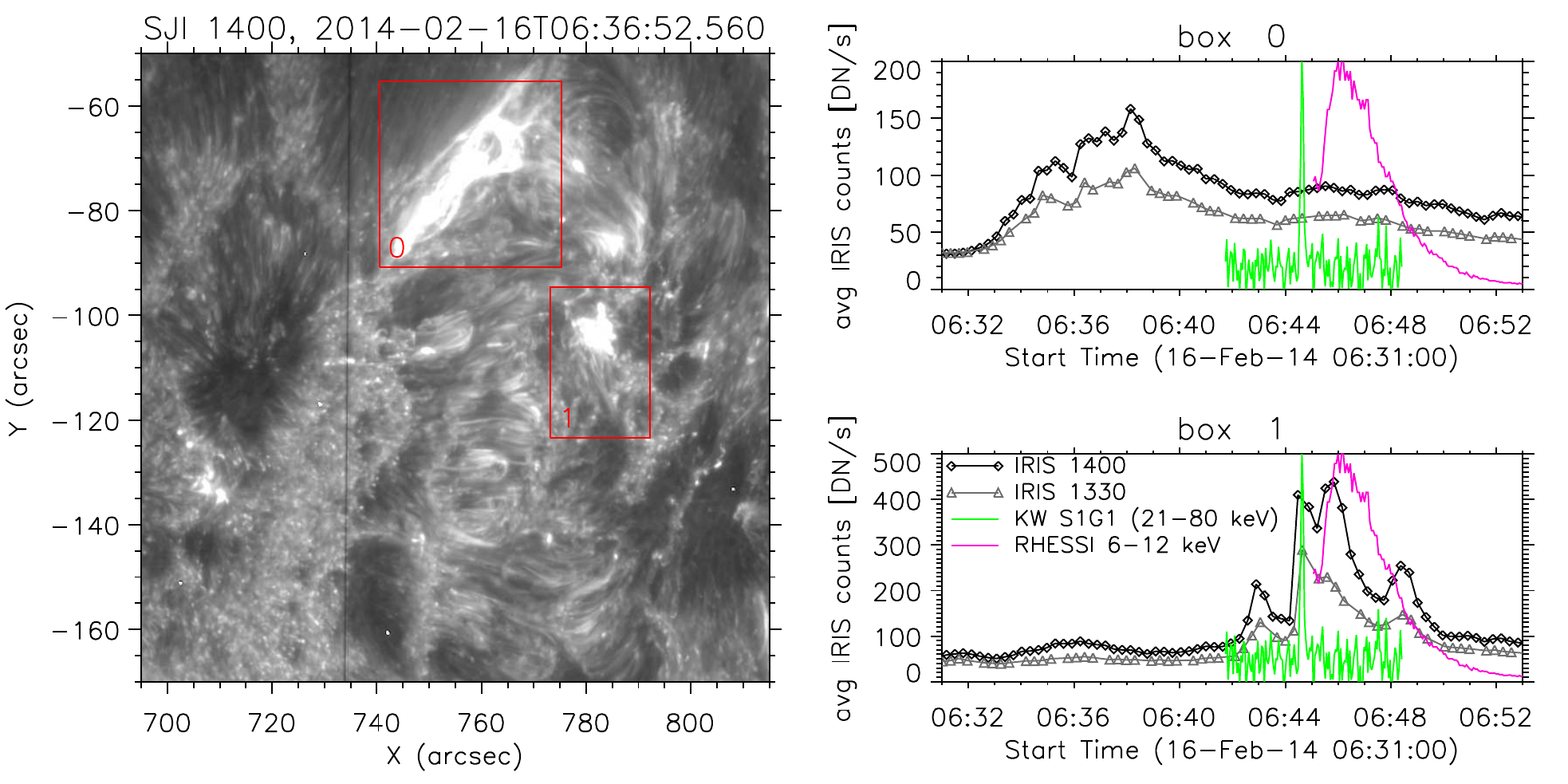}
\caption{Temporal evolution of \IRIS\ intensities. The red boxes in the left image show the areas whose \IRIS\ SJ intensities were averaged to produce the lightcurves in the right plots. It is visible that the northern loop brightened in both 1330 \AA\ and 1400\AA\ around 06:38 UT, when RHESSI was still in orbital night (and GOES did not show any significant enhancement), while the southern loop brightened around 6:45 UT coincident with the emission from other wavelengths. The \rhessi\ and \kw\ data are integrated over the whole solar disk and are shown normalized to the plot window.
\label{irislightcurves}
}
\end{figure*}

\goes\ soft X-ray (SXR) data are shown in Figure\,\ref{Fig_lightcurves}(b).  There is a preflare enhancement well seen in both low and high energy channels at $\sim$06:43\,UT,
which are roughly co-temporal with \IRIS\ enhancement from box\,1, shown in Figure\,\ref{irislightcurves}, in 1400\,\AA\ and 1330\,\AA\ (see below for \IRIS\ data) and with \sdo/AIA enhancement, seen in time profiles in Figure\,\ref{Fig_lightcurves}(c) (see next section for \sdo/AIA data analysis).

\subsection{EUV: \sdo/AIA data}

The standard EUV data set of the full solar disk is available from six \sdo/AIA (94, 131, 171, 193, 211, 335 \AA) coronal passbands. The AIA images with $\sim$1.2" spatial resolution have been taken
with 12\,s cadence, calibrated using the aia\_prep.pro routine and normalized by the exposure time. We focus on the EUV emission from the ROI shown in Figure~\ref{Fig_AIA_maps} to quantify the thermal energy and its evolution in the coronal part of the flare.
We employ the Differential Emission Measure (DEM) analysis technique applied to the entire ROI as well as using the DEM maps with the methodology developed and applied by \citet[hereafter Paper I]{2020ApJ...890...75M} to a nonthermally dominated SOL2013-11-05T035054 solar flare.
Some of the EUV images contain saturation artifacts.
Unsaturated images with shorter exposure time
are taken for the quantitative analysis of thermal energy of the flare.

%
%

\subsection{UV: \IRIS\ data}

The \IRIS\ data show activity (impulsive enhancements in a localized area) temporally coinciding with impulsive emission from \kw\ and microwaves. This is visible in the slitjaw images with passbands around 1330 \AA\ and 1400 \AA, which were obtained at a cadence of about 20 seconds. Figure~\ref{irislightcurves} shows that the enhancements of the more southern loop coincided temporally with the X-ray and microwave emission, while the northern loop brightened a few minutes earlier (while \rhessi\ was still in orbital night). The average intensity inside the two marked boxes is drawn in the right plots, for the \IRIS\ 1330 \AA\ and 1400 \AA\ passbands.

The spectra allow us to probe 
the bulk velocities of the chromospheric and transition region plasma. We analyzed the spectral lines \ion{Si}{4} 1394 \AA, \ion{Si}{4} 1403 \AA, \ion{O}{4} 1399 \AA, \ion{O}{4} 1401 \AA, \ion{O}{4} 1405 \AA, and \ion{Fe}{21} 1354 \AA, which form at different transition region to coronal temperatures.

We verified that these spectral lines do not show irregularities (double-peaks as for example observed in strong flares or supersonic flows) in our field of view and fitted a Gaussian function, with a central wavelength $\lambda_0$ and FWHM $\Delta\lambda$, to each pixel for each spectral line to determine their Doppler velocities and Doppler widths and thus to quantify the kinetic energy of the plasma flows and turbulent motions.



\subsection{MW data}

As mentioned, there are no \mw\ imaging data for this event. Only total power (spatially integrated) spectroscopic data are available from several instruments, including NoRP (with a significant flux at 3.75 and 9.4\,GHz and a weak signal at 17 GHz), RSTN (at 2.8, 5, 8.8, and 15.4\,GHz), and BBMS data at 4--8\,GHz. These data were combined in a single synthetic dynamic spectrum (see Fig.\,\ref{Fig_lightcurves}) as described in \citet{2018ApJ...856..111L}. The dynamic spectrum and single-frequency light curves are employed in conjunction with the \kw\ light curve and the 3D model of the flare to constrain the nonthermal energy deposition in this flare. 


%

\section{Thermal plasma}

\subsection{Thermal plasma diagnostics with \rhessi}\label{S_rh_diagn}

Given that \rhessi\ data are only available after the \kw\ impulsive peak is over, our expectation is to employ
\rhessi\ spectroscopy 
to quantify the hottest thermal component of the flaring plasma using OSPEX.\footnote{For documentation see \url{https://hesperia.gsfc.nasa.gov/rhessi3/software/spectroscopy/spectral-analysis-software/index.html}}
The spectral fits were applied to the background-subtracted data (detectors 1, 4, 5) every 8 s from 06:45:06~UT to 06:46:18~UT, every 12 s for the interval 06:46:18-06:47:30~UT, and every 20 s for the 06:47:30-06:53:10~UT time interval for better statistics (the intervals are respectively highlighted by light, dark, and darkest gray areas in Figure\,\ref{Fig_lightcurves}(a)).

We attempted a  two-temperature fit in the 3--17\,keV range, which returned a cool ($T\sim10$\,MK) and a hotter ($T\sim25$\,MK) component. However, uncertainties of the cool component parameters are very large; likely, because \rhessi\ calibration is unreliable below $\sim 6$\,keV. In addition, the $\chi^2$ metrics were unacceptably high due mainly to high residuals in the  $3-7$\,keV range. Thus, we only employ the hot component parameters ($T2_{\rhessi}$ and $EM2_{\rhessi}$) of this two-temperature fit, which are shown in dark green in Figure\,\ref{Fig_EM_T}.

\begin{figure}\centering
\includegraphics[width=8cm]{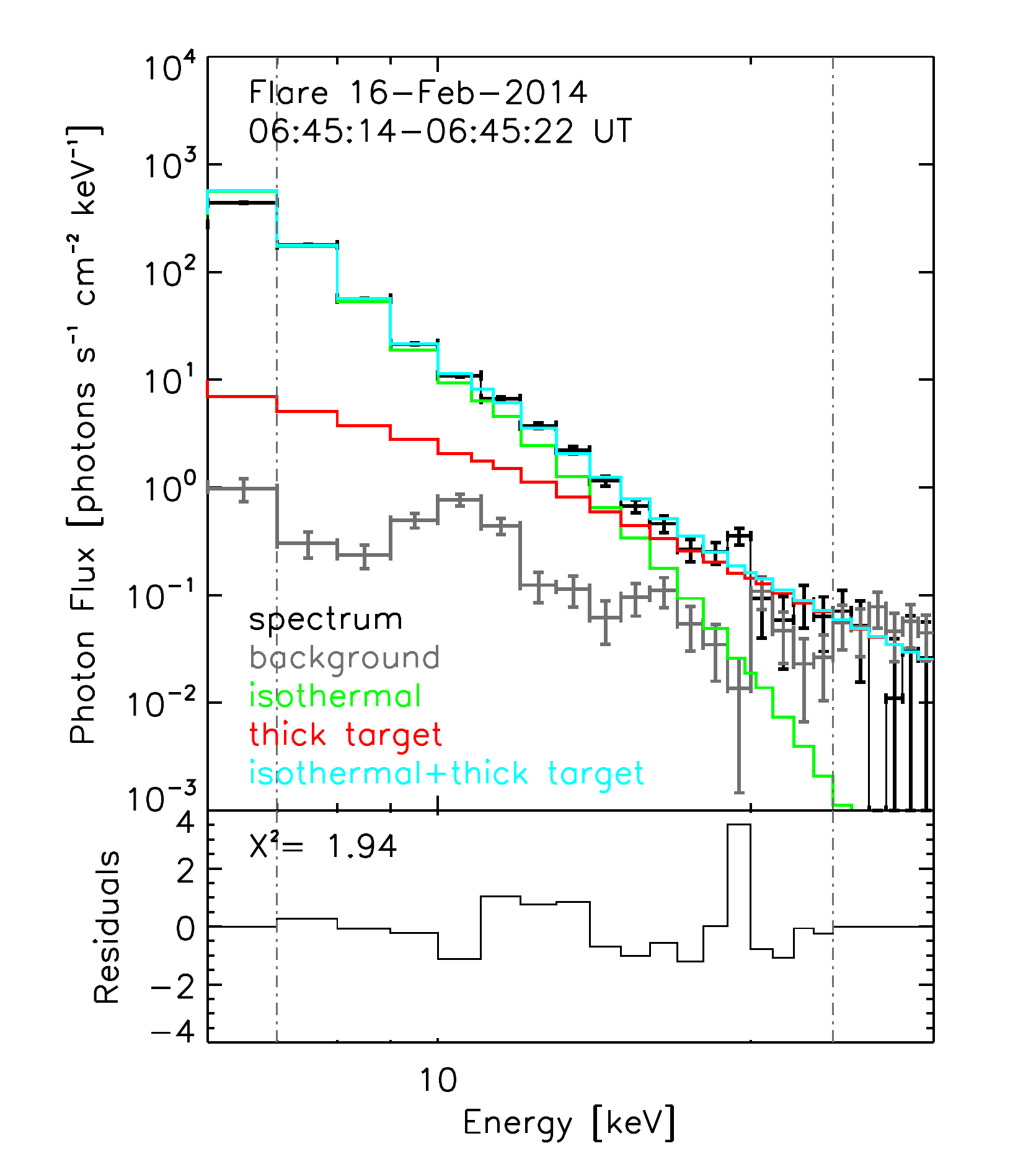}
\includegraphics[width=8cm]{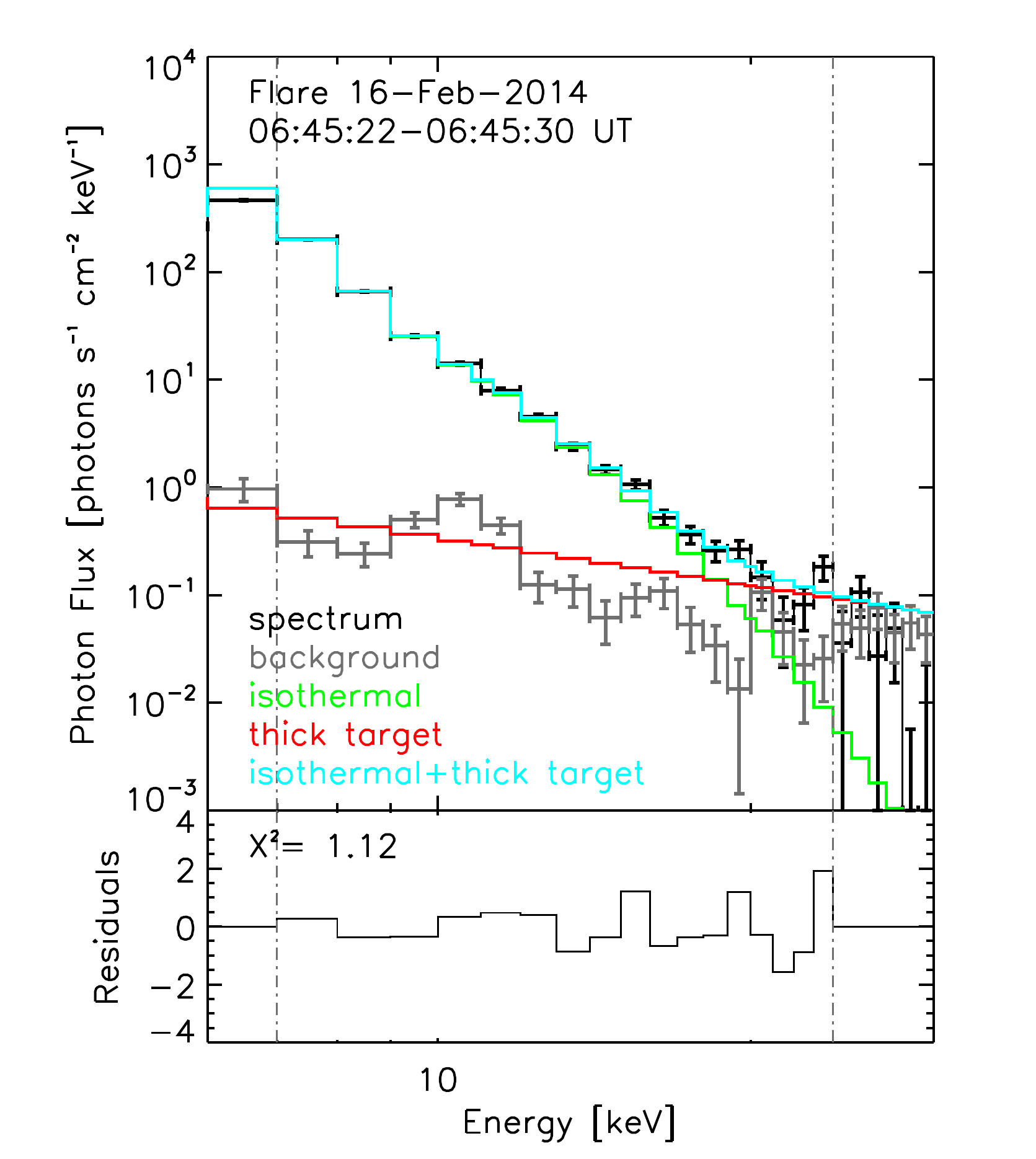}
\caption{
{Two examples (06:45:14-06:45:22~UT, top panel, and 06:45:22-06:45:30~UT, bottom panel) of \rhessi\ data (in black) and fits (in light blue). The background is shown in gray. The fit components are color coded as indicated in the panels. The residuals are shown in bottom panels of these two plots.}
\label{Fig_RHESSI_fits}
}
\end{figure}

To cross-check these results, we attempted an
isothermal plus thick-target spectral model fit in the  $7-25$\,keV range, which resulted in a better $\chi^2$ metrics  ($\chi^2<3$).  Two examples of the \rhessi\ fits, shown in Figure\,\ref{Fig_RHESSI_fits}, indicate that the nonthermal component is either very weak or nonexistent. 
The time evolution of the fit parameters is shown in Figure \ref{Fig_EM_T}. The thermal part of the fit resulted in the well-constrained  temperature (green histogram in panel a) and  emission measure (green histogram in panel b) consistent with that derived from the two-temperature fit (dark green lines).

The nonthermal component, even though appeared to be needed to return acceptably good fits, has large uncertainties of  the total integrated electron flux $F_0$ (panel c), the low energy cutoff $E_c$ (panel d), and the spectral index $\delta$ of the electron distribution function above $E_c$ (panel e). These figures confirm that there is a marginal, if any, nonthermal component in the \rhessi\ data.

The \rhessi\ spectral model fits are used to quantify the thermal energy of the hottest component of the flaring plasma similarly to the estimate reported in Paper I. 
The only distinction is that now we cannot estimate the hot loop volume from data directly. Instead, we rely on the loop volume (Loop\,II, $V_{II}=7.75\times10^{26}$\,cm$^{-3}$, see Table\,\ref{table_model_summary}) determined from 3D model devised in Section\,\ref{S_Model_Feb_16}.


\subsection{Thermal plasma diagnostics with \sdo/AIA}\label{S_aia_diagn}

\begin{figure}\centering
\includegraphics[width=0.49\columnwidth]{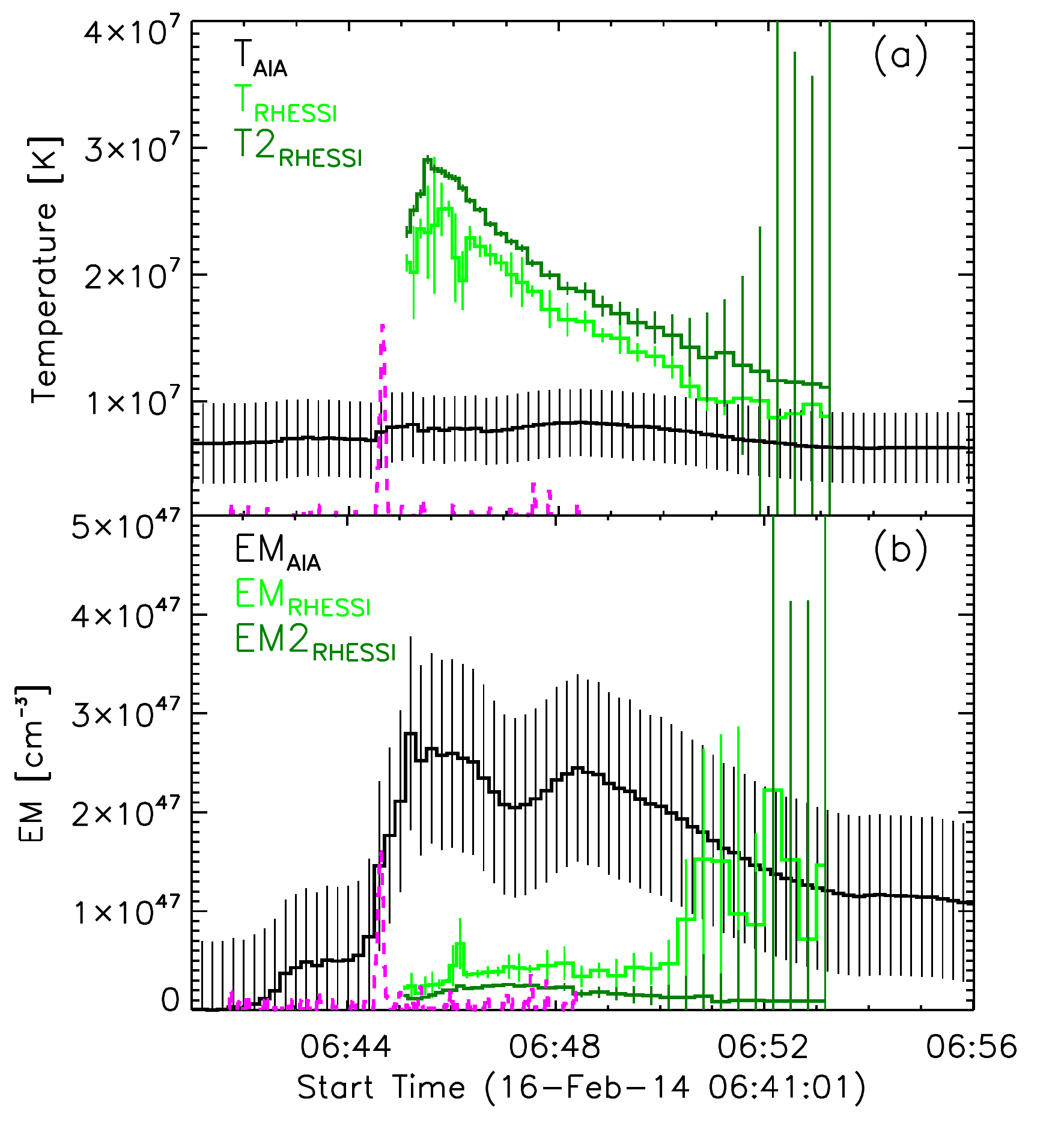}
\includegraphics[width=0.49\columnwidth]{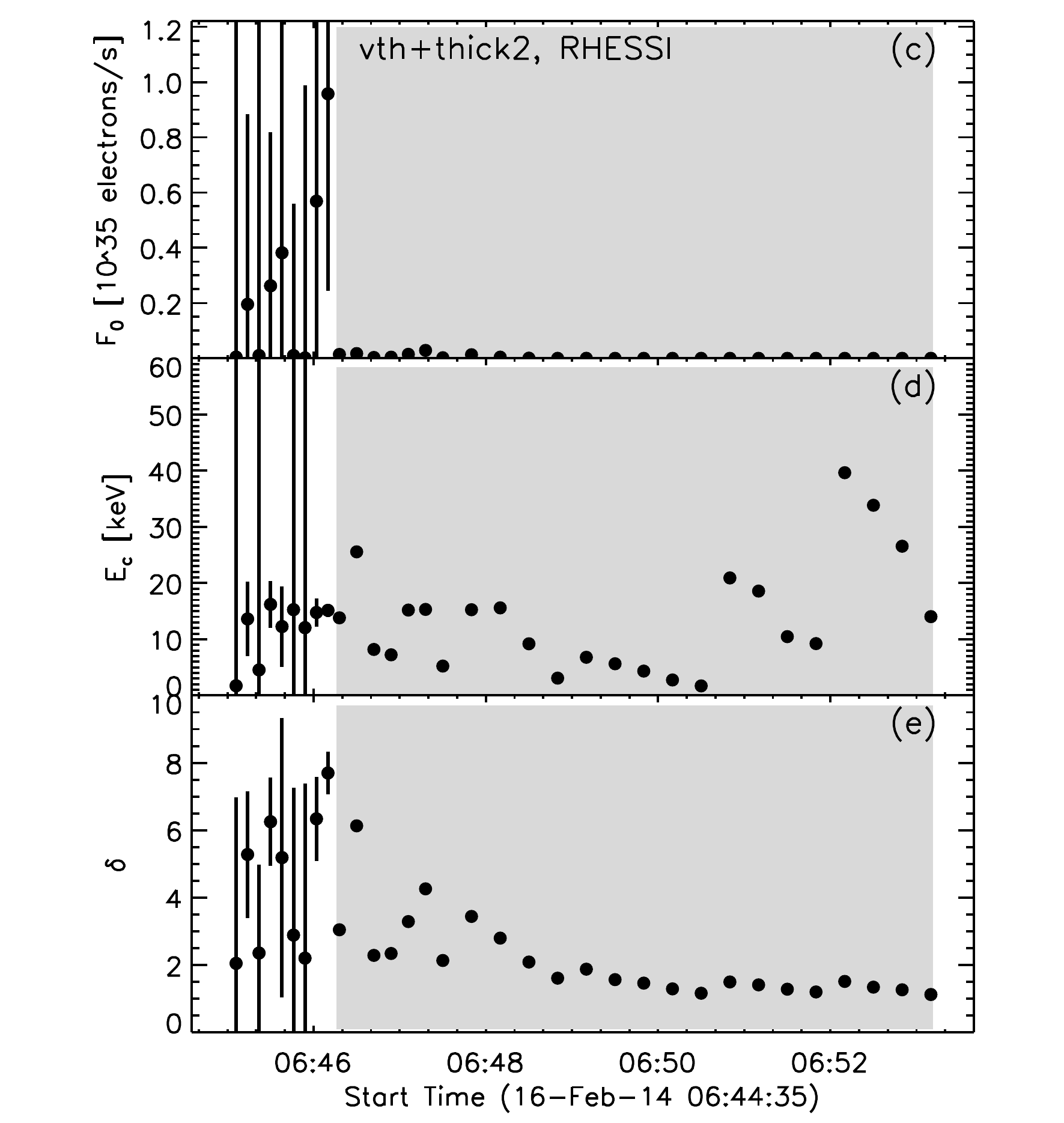}
\caption{Temporal evolution of $T$ {(a)}, $EM$ {(b), and nonthermal parameters (c, d, e)} for the February 16, 2014 flare. {(a)} \rhessi\ temperature of the hot flare component (green and dark green histograms) from two spectral fits (see text) of the spatially unresolved full solar disk and AIA temperature of the cooler flare component (black histogram) from the ROI shown in Fig.\,\ref{Fig_AIA_maps}.  {(b)} \rhessi\ (green and dark green histograms) and preflare-subtracted AIA (black histogram) emission measure, inferred as described for the upper panel. 
Red dashed histogram shows the \kw\ 21-80~keV light curve [arb. units].
(c) Total integrated electron flux $F_0$, (d) low-energy cut-off $E_c$, and
(e) spectral index $\delta$ obtained from the \rhessi\ isothermal+thick target fit. Vertical lines indicate the range of 1$\sigma$ error on the fits of the \rhessi\  data. The grayed out area indicates the time range when the formal 1$\sigma$ errors exceed the plot ranges.}
\label{Fig_EM_T}
\end{figure}

\sdo/AIA observations are used to characterise plasma
at temperatures 0.5-25 MK. To infer emission measure and temperature, we first calculate the differential emission measure (DEM) [cm$^{-5}$ K$^{-1}$] using a regularization technique \citep[e.g.][]{ti63}.
The total emission measure $EM_{\rm{AIA}}$ [cm$^{-3}$] and mean temperature $\langle T_{\rm{AIA}} \rangle$ obtained from DEMs\footnote{{The solar plasma DEM analysis employs the CHIANTI database that assumes the ionization equilibrium.
The assumption might be incorrect in case of the impulsive phase of the flare, when plasma is transitioning to equilibrium. The coronal elemental abundances are used by default. This might be incorrect if evaporated chromospheric plasma dominates the flare thermal response.}} calculated from the ROI (see Equations (1-2), Paper I) are shown in Figure\,\ref{Fig_EM_T}. The saturation effects, including a secondary saturation (blooming) are within the errors. Figure\,\ref{Fig_EM_T} shows the evolution of $EM_{\rm{AIA}}$ and $\langle T_{\rm{AIA}} \rangle$ with the minimum preflare emission measure subtracted. After the HXR impulsive peak, shown by \kw\ light curve, both the emission measure and temperature increase. Similar to 5-Nov-2013 solar flare studied in Paper\,I, the values obtained from \rhessi\ and \sdo/AIA are different from each other, which could mean that the two instruments see different sources with different temperatures. 

\begin{figure}\centering
\includegraphics[width=0.95\columnwidth]{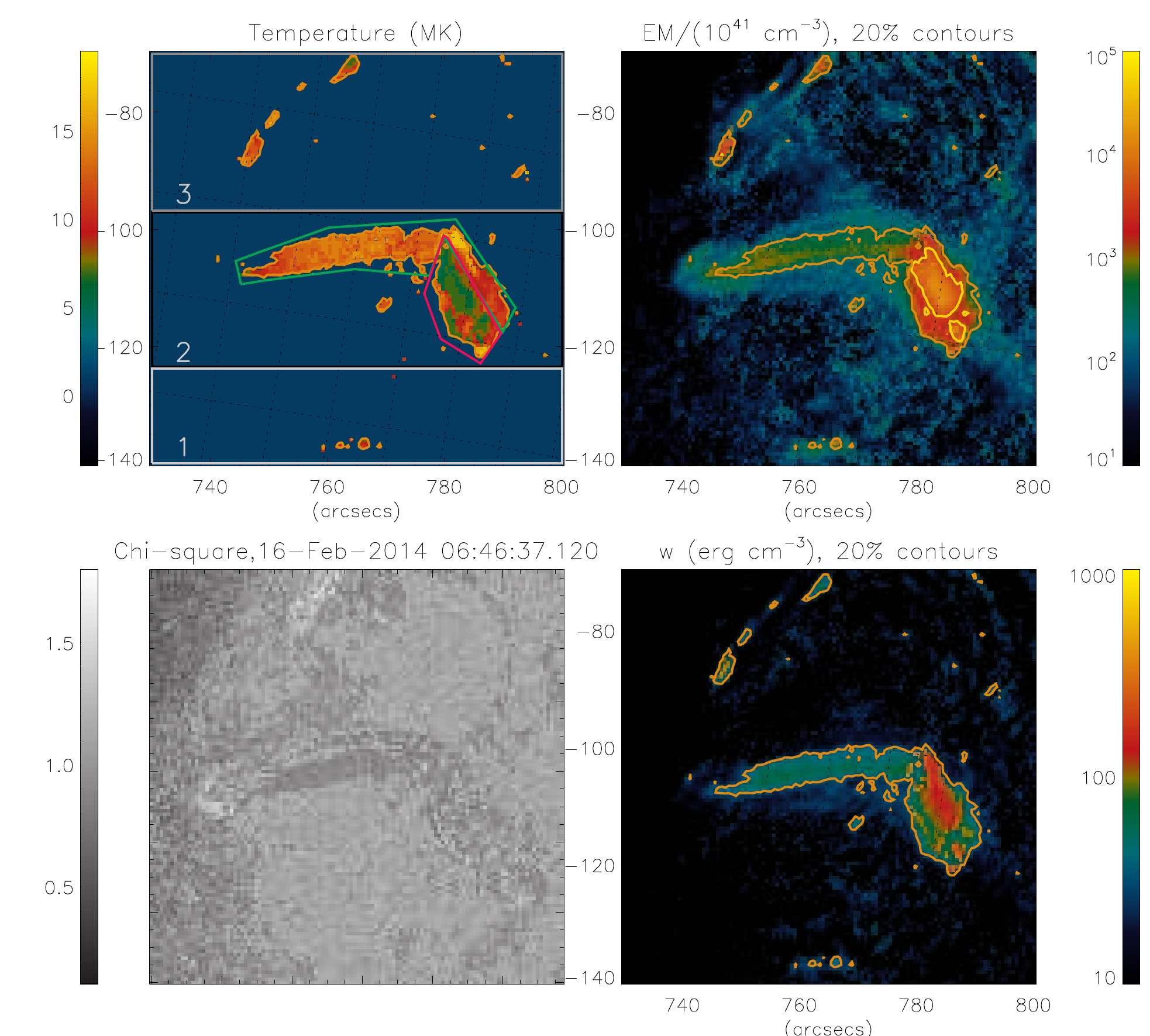}
\caption{
{Spatial distributions of plasma parameters derived from \sdo/AIA data with the regularized DEM inversion technique. The temperature map is shown in the top left panel; the emission measure map is in the top right panel; the chi-square map is in the bottom left panel; and the thermal energy density map is in bottom right panel. This Figure represents the 29th time interval (06:46:37-06:46:49~UT) of the animation that shows the entire evolution of the flare thermal parameters. To aid the eye, in each time frame we plot the contours that indicate 20\% of the thermal energy density peak (in  orange) and 20\%  of the emission measure peak (in yellow). The temperature map is plotted only within the orange contour. An animation is included with this figure. The video shows the entire flare event starting on 2014 February 16 at 06:41:01\, UT and ending the same day at 06:56:01\,UT. The video duration is 19\,s. 
}
\label{Fig_EM_map}
}
\end{figure}

\begin{figure*}\centering
\includegraphics[width=0.49\textwidth]{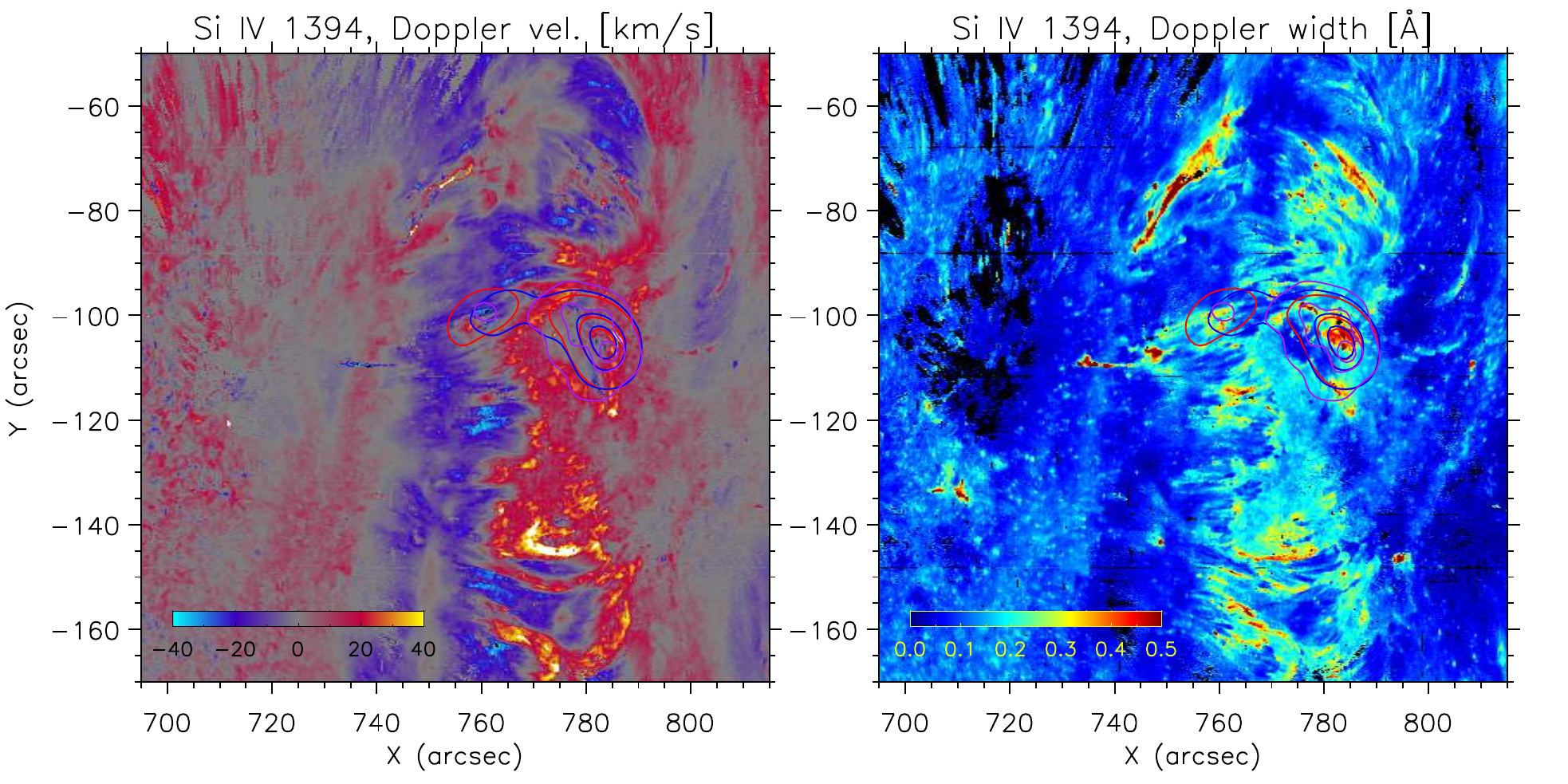}
\includegraphics[width=0.49\textwidth]{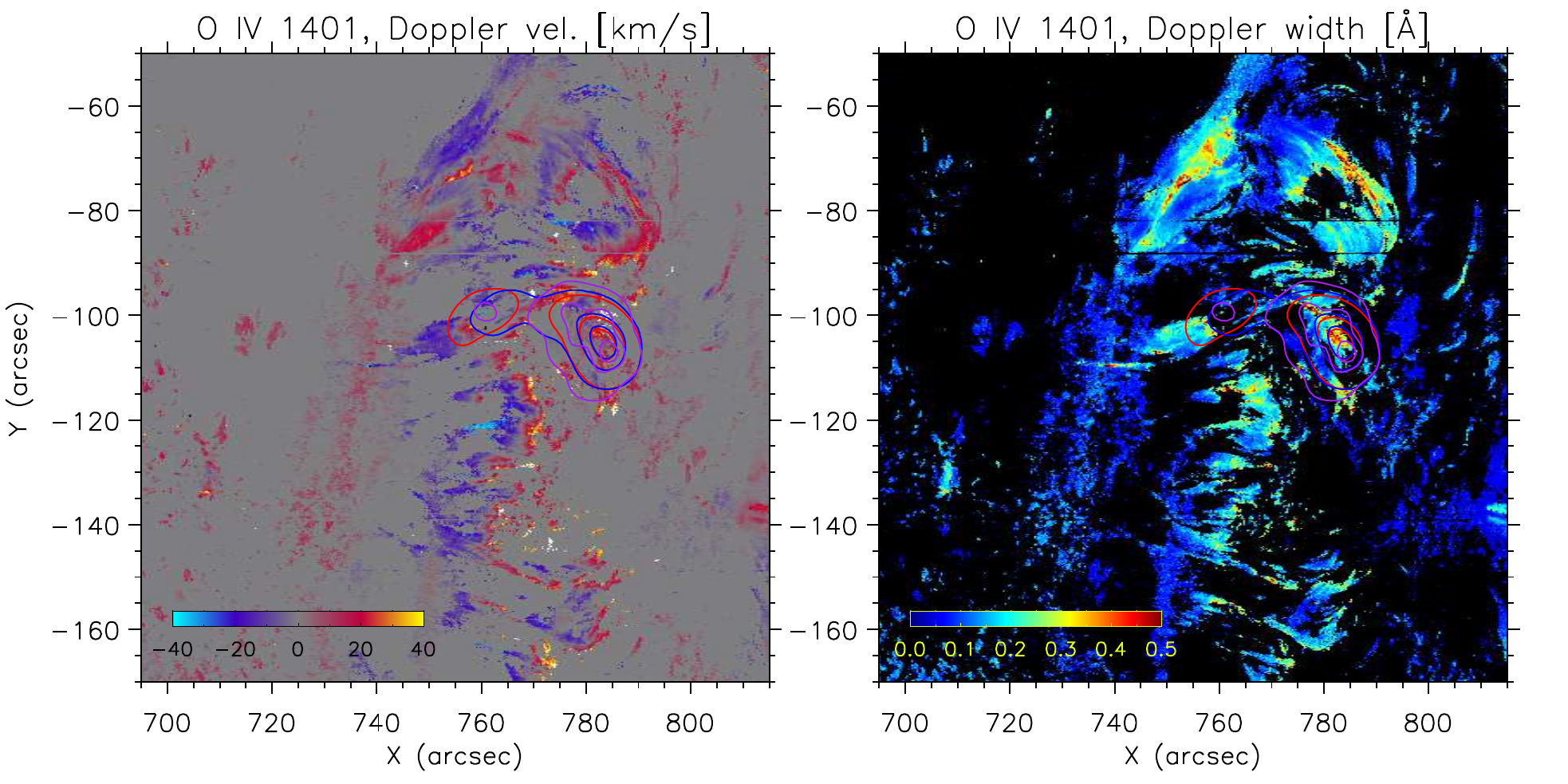}
\caption{Left panel: Doppler velocities, obtained from Gaussian fitting to the \ion{Si}{4} 1394 \AA\ line. Bad pixels (noise below cutoff, cosmic rays, ...) were set to zero velocity. Negative (blue) values indicate the blue shift (towards the Sun; positive (red) values indicate the red shift (outwards the Sun). 2nd panel: FWHM of the Gaussian, representing the Doppler widths. Bad pixels are shown in black. The flare locations show enhanced Doppler widths, but no particularly unusual Doppler velocities. 3rd and 4th panels: similar plots for \ion{O}{4} where many more pixels are below the noise limit. The pixels above the noise limit show similar properties as the corresponding ones in \ion{Si}{4} 1394 \AA.
\label{sivdoppler}
}
\end{figure*}

Applying the regularized inversion code to the \sdo/AIA data \citep{2012A&A...539A.146H,2013A&A...553A..10H} the DEM maps have been created using the methodology described in Section 2.3.2 of Paper I. The DEM maps are then used to calculate the emission measure maps $EM^{\rm{AIA}}_{ij}$ [cm$^{-3}$], the mean temperature maps $\langle T^{\rm{AIA}}_{ij} \rangle$ [K], and the  thermal energy density maps $w^{\rm{AIA}}_{ij}$ [erg cm$^{-3}$] for all $(i, j)$ pixels. 

The 3D modeling of the flare described in Section \ref{S_Model_Feb_16} shows that the essential contribution to the EUV images comes from relatively cold and dense Loop I (see Fig.\,\ref{Fig_model_0} Table \ref{table_model_summary}). Thus to estimate the thermal energy density $w^{\rm{AIA}}_{ij}$ the adopted length along the line of sight (LOS) has been taken as a characteristic diameter of the modeled Loop I (see Table\,\ref{table_model_summary}), which is
$l_{\rm{depth}}=2r \approx  3.8\times 10^8$ [cm]. 
The values $EM^{\rm{AIA}}_{ij}$, $\langle T^{\rm{AIA}}_{ij} \rangle$, $w^{\rm{AIA}}_{ij}$ are calculated using Equations (3--5) (Paper I), and are demonstrated along with the chi-square
($\chi^2$ < 2, bottom left panel) in animated Figure \ref{Fig_EM_map}. Figure \ref{Fig_EM_map} shows the 29th time interval (06:46:37-06:46:49~UT) of the animation.
The mean temperature is close to 12-14 MK during the flare, while the emission measure and the thermal energy density evolve dramatically. The saturation effects become visible after 06:45:01~UT just after the nonthermal HXR peak time. A distinct boundary between two temperatures ($\sim$8 and $\sim$12 MK) is seen in $\langle T^{\rm{AIA}}_{ij} \rangle$ maps. We interpret this as a  projection effect when due to complexity of the flare loop structure,  hot Loop II (reddish on  Figure\,\ref{Fig_EM_map}, top left panel) projects on  cooler Loop I (greenish). Note that the (pre)heating takes place in \textit{both} loops (I and II) \textit{before} the impulsive energy release, seen in the \mw\ and \kw\ data.

\subsection{Thermal plasma diagnostics with \IRIS}\label{S_iris_therm_diagn}

The coronal \ion{Fe}{21} line is sensitive to plasma temperatures around 10$^7$ K. Signatures of \ion{Fe}{21} are seen in flare loops, indicating that hot plasma.
However, these signatures are  weak (maximum of 1-2 DN/s, compared to $>$10 DN/s for other spectral lines), which precludes a reliable Gaussian fitting and thus a more quantitative analysis. {This also precludes the use of this spectral line to estimate a coronal portion of the kinetic energy.}



\section{Plasma motions in the flare}
\label{S_iris_kin_diagn}

To quantify regular and random plasma motions 
of the flare, we focus on the spectral lines of \ion{Si}{4} (1393.76 and 1402.77 \AA) and \ion{O}{4} (1399.78, 1401.16, and 1404.78 \AA), which form at temperatures of 10$^{4.8}$ and 10$^{5.1}$ K, respectively \citep{2014iris,youngetal2018}.
{Note that this temperature range pertains to the flare footpoints, {and} does not capture the coronal portion of the flare. Thus, these data quantify only a fraction of the total kinetic energy in the flare.}

\subsection{Bulk plasma velocity}

Gaussian fitting allows us to determine the displacement of the Gaussian peak value $\lambda_0$ from the expected rest frame wavelength  $\lambda_{ji}$ of the transition between the levels $i$ and $j$, which yields the bulk velocity due to the Doppler effect:
\begin{equation}
\label{Eq_doppler_shift}
\frac{v_{bulk}}{c}= \frac{\lambda_0-\lambda_{ji}}{\lambda_{ji}},
\end{equation}
where $c$ denotes the speed of light. 

The Doppler fitting of \ion{Si}{4} 1394 is very reliable, with few exclusions necessary: about 5\% of the pixels were below our selected cutoff for the noise (1.25 DN/s) and were therefore excluded. Furthermore, we excluded pixels with cosmic rays, spikes, or defects (based on the fitted FWHM $\Delta\lambda$), amounting to 2\% of the FOV.

Figure~\ref{sivdoppler} shows enhanced Doppler widths at the location of the \rhessi\ source in \ion{Si}{4} 1394 \AA\ and also in the more northern loop (0.5 \AA\ vs.\,0.2-0.3 \AA\ in the quiet Sun). Nearly identical values were obtained from \ion{Si}{4} 1403 \AA\ (therefore not shown). The \ion{Si}{4} Doppler velocities were not very noteworthy and ranged between $\pm$ 20 km s$^{-1}$, which is a common range also in the quiet Sun as shown in the left panel. Excluded pixels are shown in black for the Doppler widths and with a value of zero (gray) for the Doppler velocity.

\subsection{Turbulent plasma velocity}
\label{S_iris_turb_diagn}


{The total line broadening depends on three quantities: the thermal line width $ \frac{2 k_B T_{ion}}{M_{ion}}$ [km s$^{-1}$], the instrumental broadening ($\sigma_I$) [km s$^{-1}$], and the non-thermal line width ($v_{turb}$) [km s$^{-1}$]. Thus, the FWHM $\Delta\lambda$ of a spectral line in units of \AA\ can be written as}

\begin{equation}
\label{Eq_doppler_width}
\Delta\lambda = \frac{\lambda_{0}}{c} \sqrt{4 \ln2 \left( \frac{2k_B T_{ion}}{M_{ion}}
+ v_{turb}^2 + \sigma_I^2
\right)}.
\end{equation}
{
The \IRIS\ instrumental broadening was determined both by lab measurements \citep{tianetal2014} and by measuring the \ion{O}{1} 1355.598 \AA\ line \citep{2014iris}, and yielded for the spectral range around the \ion{Si}{4} line about 30 m\AA\ (6.4 km s$^{-1}$). The thermal broadening for \ion{Si}{4} is 50 m\AA\ ($\approx 11$ km\,s$^{-1}$), and for \ion{O}{4} 90 m\AA\ ($\approx 19$ km\, s$^{-1}$).\footnote{\url{https://iris.lmsal.com/itn38/}}
}

{Because of the better signal to noise ratio, we use the \ion{Si}{4} line to calculate the non-thermal line width based on its Gaussian fits, which allowed us to determine the total FWHM and solve Eq.~\ref{Eq_doppler_width} for $v_{turb}$. We caution that this is only an approximation because we derive the FWHM by fitting single Gaussian, which may not be entirely accurate in some flare pixels \citep{2016A&A...590A..99J}.
A visual inspection of the spectra shows that this seems reasonable in most pixels, but there are pixels with broad asymmetric spectra, whose origin probably is a superposition of multiple atmospheric components, for example multiple downflows of different velocities within one pixel. Such pixels lead to an overestimation of the turbulent velocities and therefore our number given here is rather an upper limit.}



\subsection{Density diagnostics using \ion{O}{4} lines}

Several \ion{O}{4} lines are observed in \IRIS' spectrum. They are generally weak, but are visible in some locations, particularly around our flare loops. We carried out a similar Doppler fitting of the \ion{O}{4} 1401.156 \AA\ line and in this case, 73\% of all pixels had to be excluded due to being too noisy and an additional 5\% because of cosmic rays or the fitted FWHM being below our cutoff (50 m\AA). Nevertheless, the good pixels appeared to have very similar properties to \ion{Si}{4} 1394 \AA. We therefore later use the \ion{Si}{4} 1394 \AA\ Doppler velocity as a proxy for the \ion{O}{4} 1401.156 \AA\ Doppler velocity.

\begin{figure}\centering
\includegraphics[width=0.49\textwidth]{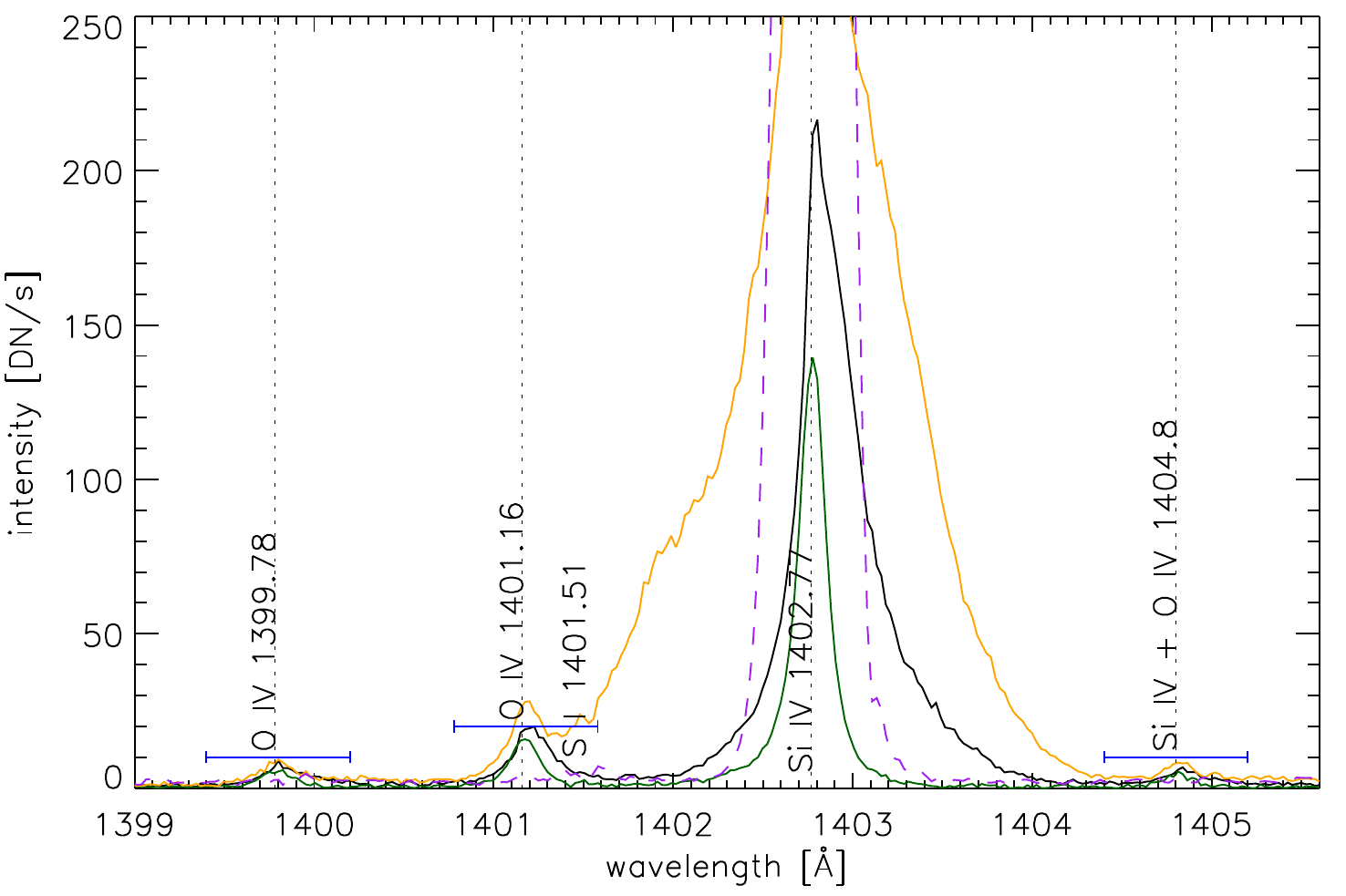}
\caption{
Spectra of example pixels (different colors) showing the \ion{O}{4} lines near 1400 \AA. The blue bars indicate the wavelength range over which the intensity was integrated to determine the total line intensity used to calculate the line ratios. It is visible that the \ion{O}{4} lines are weak and can often be below the noise level. In case of large velocity flows, the 1401.16 \AA\ line can be blended by the nearby \ion{Si}{4} line, further complicating determining its integrated intensity.\label{oivexample}}
\end{figure}

\begin{figure*}\centering
\includegraphics[width=0.9\textwidth]{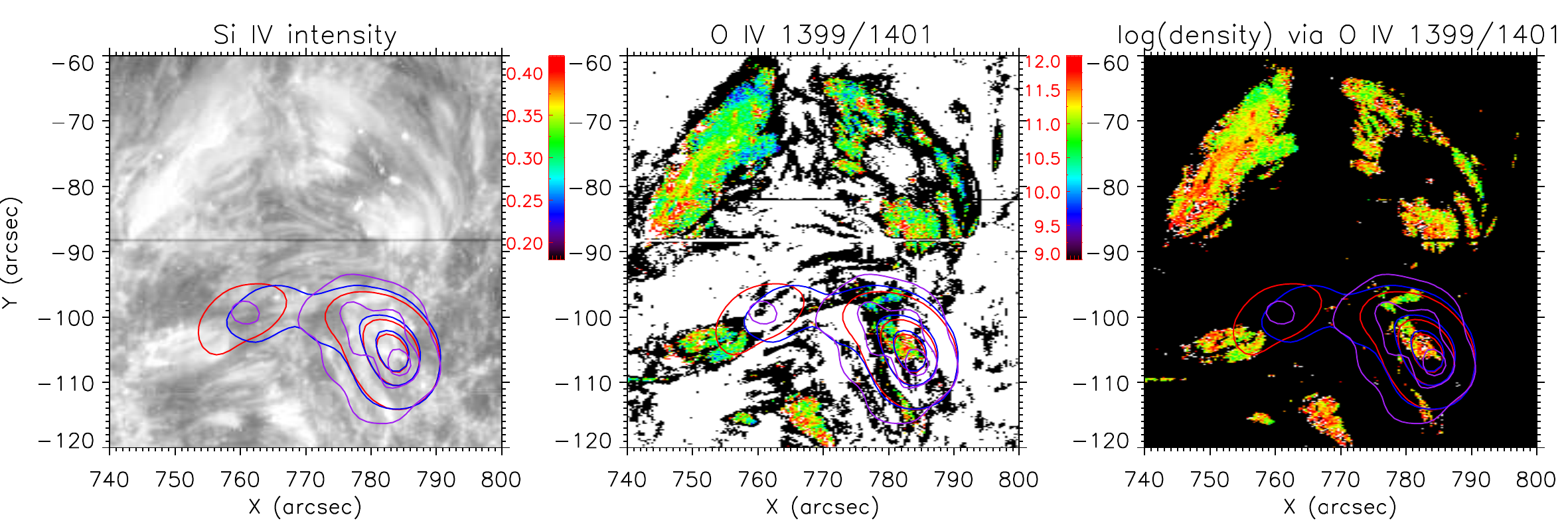}
\caption{Left: \ion{Si}{4} intensity showing the flare loops and overplotted RHESSI contours (identical color-coding to Fig.~\ref{irisoverview}). Middle panel: Ratio of integrated intensities of the \ion{O}{4} 1399.8/1401.2 \AA\ lines. White and black pixels met one of the exclusion criteria (see text). Right panel: Ratio converted to densities. The colorbar indicates the log of the densities, which can range from 10$^9$ -- 10$^{12}$ cm$^{-3}$.
\label{oivdensities}}
\end{figure*}

Line intensity ratios of several \ion{O}{4} lines around 1400\,\AA\ can be used as  diagnostics for electron densities in the range of 10$^9$ - 10$^{11}$ cm$^{-3}$ and weakly up to 10$^{12}$ cm$^{-3}$
\citep[e.g.][]{flowernussbaumer1975, doscheketal2016, politoetal2016, youngetal2018}. Our goal is to estimate the electron density near the \rhessi\ source to be able to calculate the kinetic energy density. We use the ratio of the \ion{O}{4} 1399.8/1401.2 lines because it is least blended as shown in the example spectra in Fig.~\ref{oivexample}. A drawback is the low intensity of the 1399.8 \AA\ line, which is below the noise level for many pixels. We integrate over a range of 0.8 \AA\ (indicated with blue lines in Fig.~\ref{oivexample}) and apply selection criteria on the integrated intensity and the ratios to exclude noise, cosmic rays, or unusual spectra. We visually verified that the remaining spectra show regular line profiles and produce ratios in the expected range (0.18-0.42). We 
assumed thermal and ionization equilibrium to apply this density diagnostic. Although it is difficult  to firmly justify,  considering the weakness of the flare, plus that the slit crossed it after its impulsive phase, we do not expect big errors with this assumption. Figure~\ref{oivdensities} shows our resulting ratios and densities. The ratios (middle panel) can only be determined near the flare site. Black indicates a ratio near zero, meaning that the 1399 \AA\ line is below the noise level, while white pixels indicate that one of the exclusion criteria was met, which includes too small integrated intensities, and abnormal ratios e.g.\,due to cosmic rays or large Doppler velocities and thus blending of the \ion{Si}{4} line. Even though the accuracy of this method can be debated, mostly because of the low \ion{O}{4} signals in the present observation, the figure shows that the densities of our region of interest are around 10$^{11}$--10$^{12}$ cm$^{-3}$ with few (measurable) densities below those values.





\newpage

\section{Modeling and model validation}
\label{S_Model_Feb_16}

Using the data described in the previous section, we are in the position to quantify the thermal and kinetic energies, but we cannot quantify the nonthermal energy (deposition) with the available data alone; cf. Paper\,I. Here, to estimate the nonthermal electron component in the flare, we are forced to combine the incomplete \mw\ and \kw\ data with 3D modeling of the flare based on NLFFF reconstruction. This reconstruction has to be initiated with the photospheric vector boundary condition available for this event from both \sdo/HMI and \hinode/SOT.

\subsection{Model creation via the pipeline}

In this study we employed the automated model production pipeline 
(Nita et al. 2021; in preparation)
based on the NLFFF extrapolation code \citep{2017ApJ...839...30F} initiated with an \sdo/HMI vector magnetogram taken at 06:34:12\,UT. Visual inspection of automatically downloaded base maps does not reveal any apparent inversion or $\pi$-disambiguation artifact; so the model did not require any manual correction \citep[cf.][]{2019ApJ...880L..29A}.
%
%
%

\begin{figure}\centering
\includegraphics[width=0.49\textwidth]{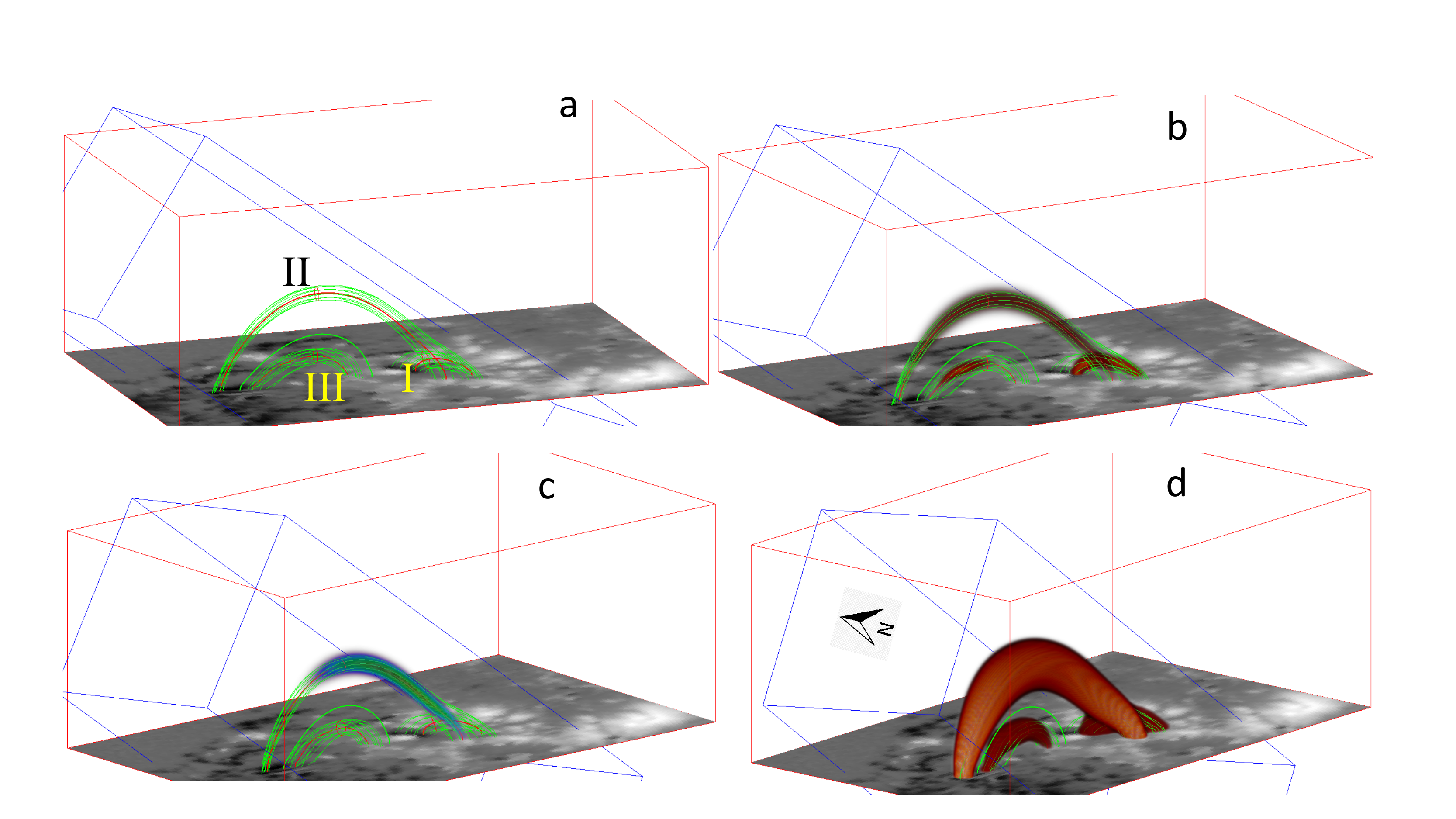} 
\caption{
The 3D model with three flux tubes (closed loops), which are labeled with their numbers (I--III) in panel a. a: magnetic flux tubes; b: distribution of thermal number density in the flux tubes; c: distribution of nonthermal number density in the flux tubes (in Loop II only); d: distribution of the temperature in the flux tubes. Red lines indicate the outer borders of magnetic data cube, where the coronal magnetic model has been computed. The bottom of the data cube shows the photospheric LOS magnetogram. The blue lines outline the scanbox from the outer (upper left) boundary of which the emissions are computed within the GX Simulator. The arrow in panel d shows the solar north direction. 
An animation is included with this figure. The video sequentially shows 360$^o$ rotations of panels (b), (c) and (d). The video duration is 9 s.
\label{Fig_model_0}
}
\end{figure}

\subsection{Selection of the flaring loops}

The NLFFF 3D magnetic data cube is imported in the modeling tool, GX Simulator \citep{2015ApJ...799..236N, 2018ApJ...853...66N}, where the flaring flux tubes are interactively created. The GX Simulator functionality permits computation (and visualisation) of selected magnetic field lines such as to match available flare images. In our case, we used only images of thermal emission obtained from \rhessi\ data at various energy ranges and also emission measure (EM) maps obtained from the DEM/AIA maps. This yielded two distinct flux tubes; Loops I and II, see Figure\,\ref{Fig_model_0} and Table\,\ref{table_model_summary}.

\begin{figure}\centering
\includegraphics[width=0.49\textwidth]{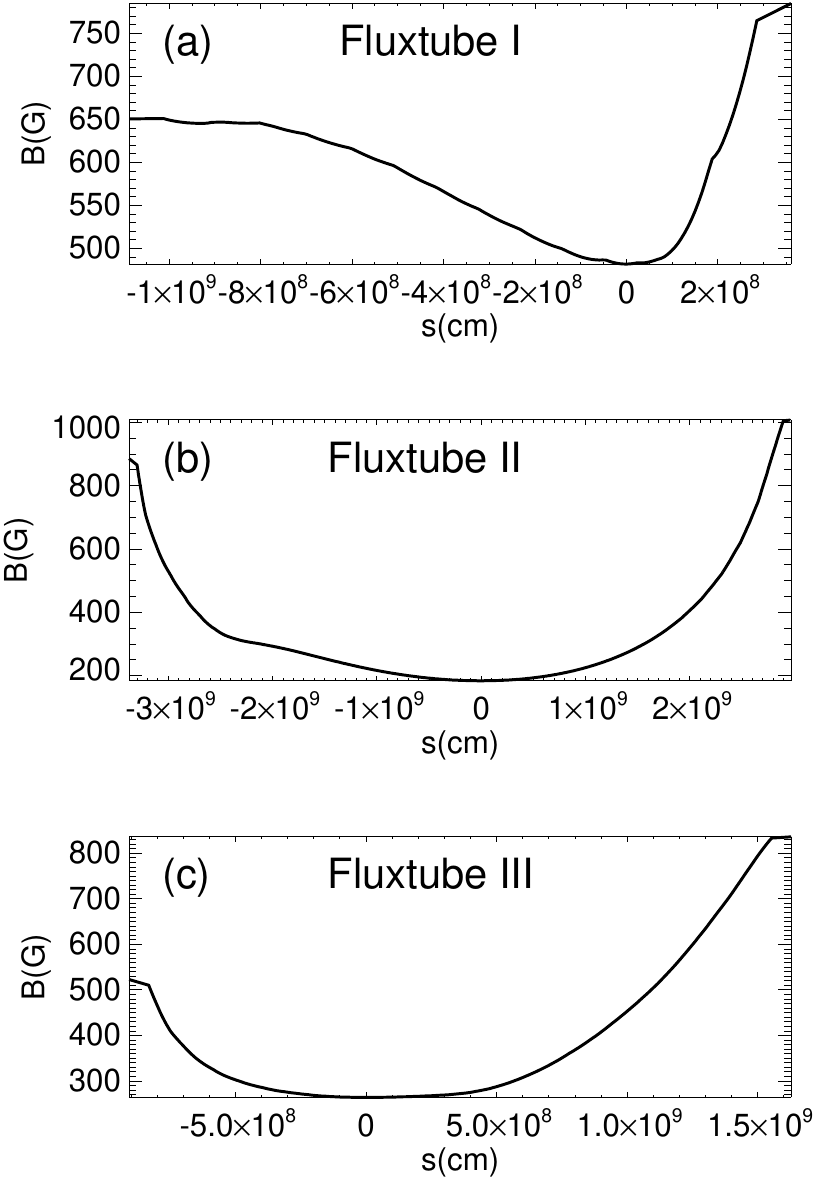} 
\caption{Distributions of the absolute value of the magnetic field along the spines of three flux tubes selected for 3D modeling of the flare.
\label{Fig_model_B}
}
\end{figure}

\begin{figure*}\centering
\includegraphics[width=.47\textwidth,angle=90]{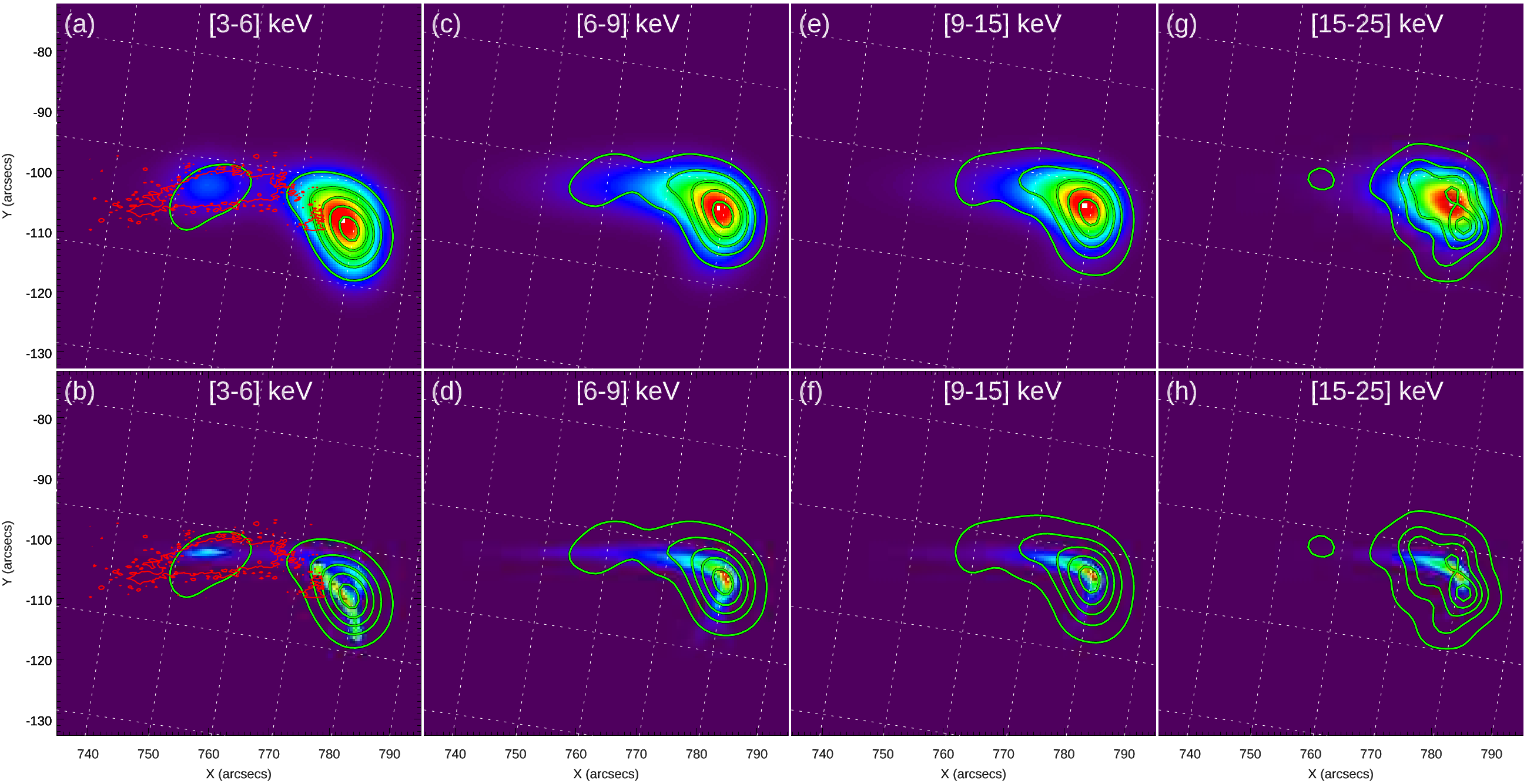}
\caption{GX Simulator-generated X-ray images from the three-loop model integrated over the ranges  3-6, 6-9, 9-15, and 15-25 keV, convolved with the \rhessi\ PSF (first row) and non-convolved (second row).  The \rhessi\ CLEAN  10, 30, 50, 70, 90\% contours (green lines) at the same energy ranges are overplotted. In addition, the integrated \IRIS\ intensities at 1353.9-1354.15\,\AA\ (\ion{Fe}{21}) are shown in panels a and b (magenta). The bright FeXXI emission comes from Loop 3 as well as the eastern \rhessi\ source at 3-6\,keV. The bottom row shows cleanly three distinct contributions from three model loops. These contributions are smeared out by the finite spatial resolution and so not distinguishable in the top row.
\rhessi\ images are synthesised for the time range 06:45:20-06:46:20\,UT. The \IRIS\ image is for the time range when the slit crossed this region, which lasted about 15 minutes.
To co-align the \rhessi\ and model maps, we applied the \rhessi\ roll angle rhessi\_roll\_angle$=-0.2$  to rhessi\_roll\_center=(0, 0), similarly as in Fig.\,\ref{Fig_AIA_maps}.
This roll angle is equivalent to $y$ shift of $-4\arcsec$. This is applied to 3 keV maps and to non-convolved maps, while $\Delta x=1\arcsec$, $\Delta y=-2\arcsec$ was applied to higher-energy convolved maps. This additional mismatch might be provided by an error of the connectivity reconstruction in the NLFFF model \citep{2019ApJ...870..101F}, or a minor inaccuracy in the spatial distribution of the the thermal plasma in loop 2 (longest, hottest one).
\label{Fig_xray_image}
}
\end{figure*}

Guided by the thermal plasma parameters derived from the OSPEX fit to the \rhessi\ data and EM maps obtained from AIA data, we succeeded to reproduce most of the imaging data, but the eastern source at the \rhessi\ 3-6\,keV image. To reproduce this source, we created one more flux tube (Loop III) filled with a relatively cool plasma (10\,MK). {The magnetic field profiles along the central (reference) field lines in all these three loops are shown in Figure\,\ref{Fig_model_B}.} With these three loops all thermal images are closely reproduced in the model; 
Figure\,\ref{Fig_xray_image}. Loop III  projects on the \IRIS\ \ion{Fe}{21} source indicative of the 10\,MK plasma there; see Figure\,\ref{Fig_xray_image}.
Figure\,\ref{Fig_xray_image} also illustrates the effect of the \rhessi\ point-spread-function (psf), which smears contributions from various thermal flux tubes out, producing a (misleading) visual impression that only one single loop is present.






\subsection{Population of the flaring loops with nonthermal electrons}
\label{S_nonthermal_population}

The next step of flare model creation is populating the flux tubes with nonthermal electron components, which ideally to be guided by images of nonthermal emission. However, such images are not available in our case; thus, we are forced to employ the only available \mw\ spectral data. To be specific, we focus on the \mw\ emission peak time at 
06:44:41\,UT.

There could be an ambiguity how exactly to distribute nonthermal electrons between three available thermal loops of our model given the absence of the relevant images. Moreover, bulk of the flaring \mw\ emission from a flare can come from entirely different flux tubes, not visible as a thermal source \citep{Fl_etal_2017, 2018ApJ...852...32K}.

Luckily, for the given flare this ambiguity can be largely removed {due to strong sensitivity of the \gs\ spectrum to the magnetic field in the emission source \citep[][Supplementary Materials including movie S2]{2020Sci...367..278F}: the stronger the magnetic field the larger the spectral peak frequency.
Combining this property of the \gs\ spectrum with dependence of its high-frequency slope on the nonthermal electron energy spectrum and of its low-frequency slope on the source geometry \citep{2020Sci...367..278F},}
we found, that populating either Loop I or III with noticeable amount of nonthermal electrons results in overestimating the \mw\ emission at high frequencies and, simultaneously, underestimating it at the low frequencies.

\begin{figure}\centering
\includegraphics[width=0.97\columnwidth]{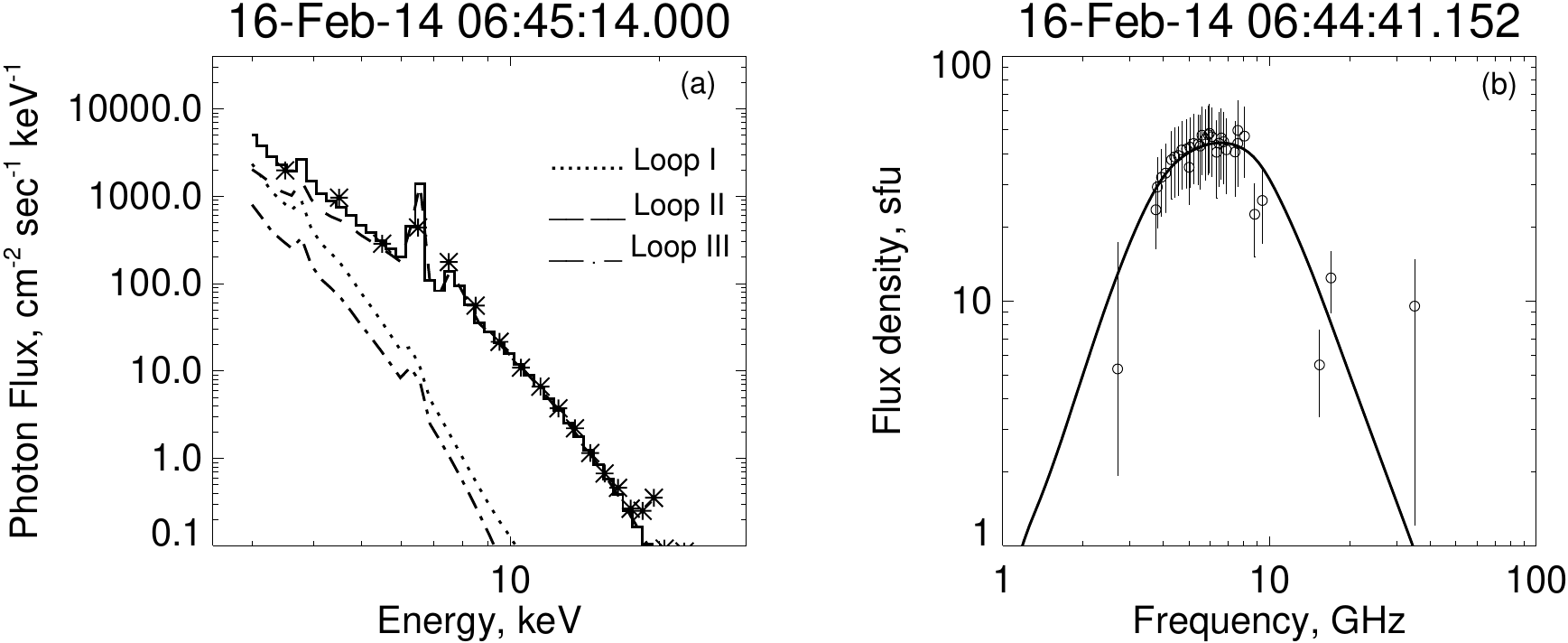}
\caption{(a) The observed {\rhessi} X-ray spectrum for time interval 06:45:14-06:45:22\,UT (asterisks) and the corresponding simulated spectrum (histogram) from the 3D model. Contributions from three distinct model loops are shown by various lines. (b) The same for the radio domain with NoRP, RSTN, and BBMS data at 04:44:41\,UT (circles) and the simulated microwave spectrum  (black solid line). Only total model spectrum is shown because contributions to the radio emission from Loops I and III are negligible. Note: the time frames selected for panels (a) and (b) are not the same because of \rhessi\ night during the impulsive flare phase.
\label{Fig_model_radio_spec}
}
\end{figure}

This conclusion appears model-independent and related only to the fact that the magnetic field is too high in those two loops: if we match the spectral peak frequency, then, the model flux level is way too low; if we match the flux level (at a high frequency), then the spectral peak frequency is much higher than observed. After careful investigation, we concluded that energetically dominant fraction of the nonthermal electrons must be located in Loop II\footnote{{It would be desirable to estimate the upper limits of the nonthermal electrons in flux tubes I and III; but it is hard to do without imaging data. The only safe statement would be that those numbers are undetectably small vs Loop II numbers.}}; see Table\,\ref{table_model_summary}, with which we can nicely reproduce the synthetic \mw\ spectrum as shown in Figure\,\ref{Fig_model_radio_spec}. The model outlined in Table\,\ref{table_model_summary} is consistent with all available observational constraints and, thus, validated by the data.

With this model we can also check if the presence of a modest nonthermal component derived from the \rhessi\ fit in Section\,\ref{S_rh_diagn} is consistent with the data. To this end, we populated flux tube 2 with nonthermal electrons consistent with the \rhessi\ fit and computed the \mw\ emission. We found that the \mw\ emission produced by this nonthermal electron population is below the 1\,sfu level at all frequencies and, thus, not observable. Therefore, we cannot confidently exclude the presence of a nonthermal electron population as derived from the \rhessi\ fit even that the uncertainties of the associated parameters are rather large.

\newpage



%
%

\begin{table*}[ht]
\caption{Summary of the 3D model}
\begin{tabular}{l l l l l}
\hline\hline
Parameter & Symbol, units &  Loop I &  Loop II &  Loop III\\ [0.5ex]
\hline
{\textit{Geometry}:} &  & \\
\quad Length of the Central Field Line     & $l$, cm  & $1.448\times10^9$ & $6.345\times10^9$ & $2.535\times10^9$  \\
\quad Reference radius of the flux tube at looptop  & $r$, cm  & $1.9\times10^8$ & $1.52\times10^8$ & $1.52\times10^8$  \\
\quad Model Volume; $\left[\int n_0 dV\right]^2/\int n_0^2 dV$ & $V$, cm$^3$ & $7.27\times10^{25}$ & $7.75\times10^{26}$ & $4.03\times10^{25}$  \\
{\textit{Thermal plasma}:} &  & & \\
\quad Emission Measure, $\int n_0^2 dV$ & $EM$,  cm$^{-3}$ & $8.91\times10^{46}$ & $1.57\times10^{46}$  & $2.58\times10^{46}$ \\
\quad Mean Number Density, $\int n_0^2 dV/\int n_0 dV$ & $n_{th}$, cm$^{-3}$ & $3.5\times10^{10}$ & $0.45\times10^{10}$& $2.53\times10^{10}$ \\
\quad Temperature      & $T$, MK  &  9 & 25  & 10 \\
\quad Instant Total Thermal Energy & $W_{th}$, erg &  $0.95\times10^{28}$ & $3.61\times10^{28}$  & $4.23\times10^{27}$\\
\textit{Nonthermal electrons:} &  & & \\
\quad Total electron number & $N_b$, cm$^{-3}$ &  $-$ & $2.41\times10^{35}$ & $-$   \\
\quad Low-/High- energy Cutoff & $E_0$, MeV & $-$  & 0.01/2 & $-$   \\
\quad Spectral Index & $\delta$ & $-$ & 3.9  & $-$\\
\quad Instant Total Nonthermal Energy & $W_{nth}$, erg &  $-$ & $5.87\times10^{27}$ & $-$ \\
[1ex]
\hline
\end{tabular}
\label{table_model_summary}
\end{table*}

\section{Energy partitions and evolution}

In this section we use the model and the data analysis products described above to quantify energies, energy partitions, and evolution thereof.

\subsection{Thermal energy derived from EUV data}

\begin{figure}\centering
\includegraphics[width=0.95\columnwidth]{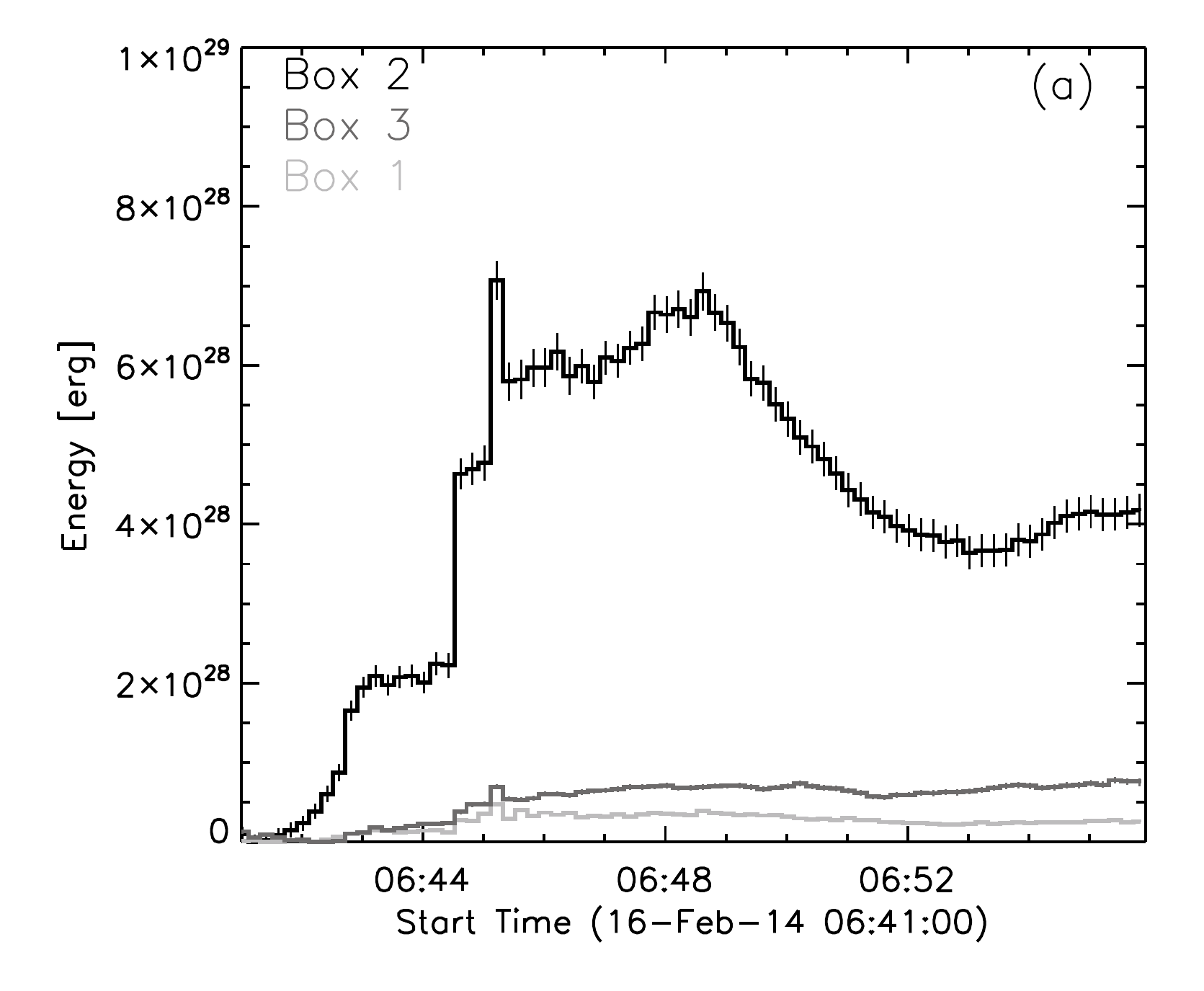}
\includegraphics[width=0.95\columnwidth]{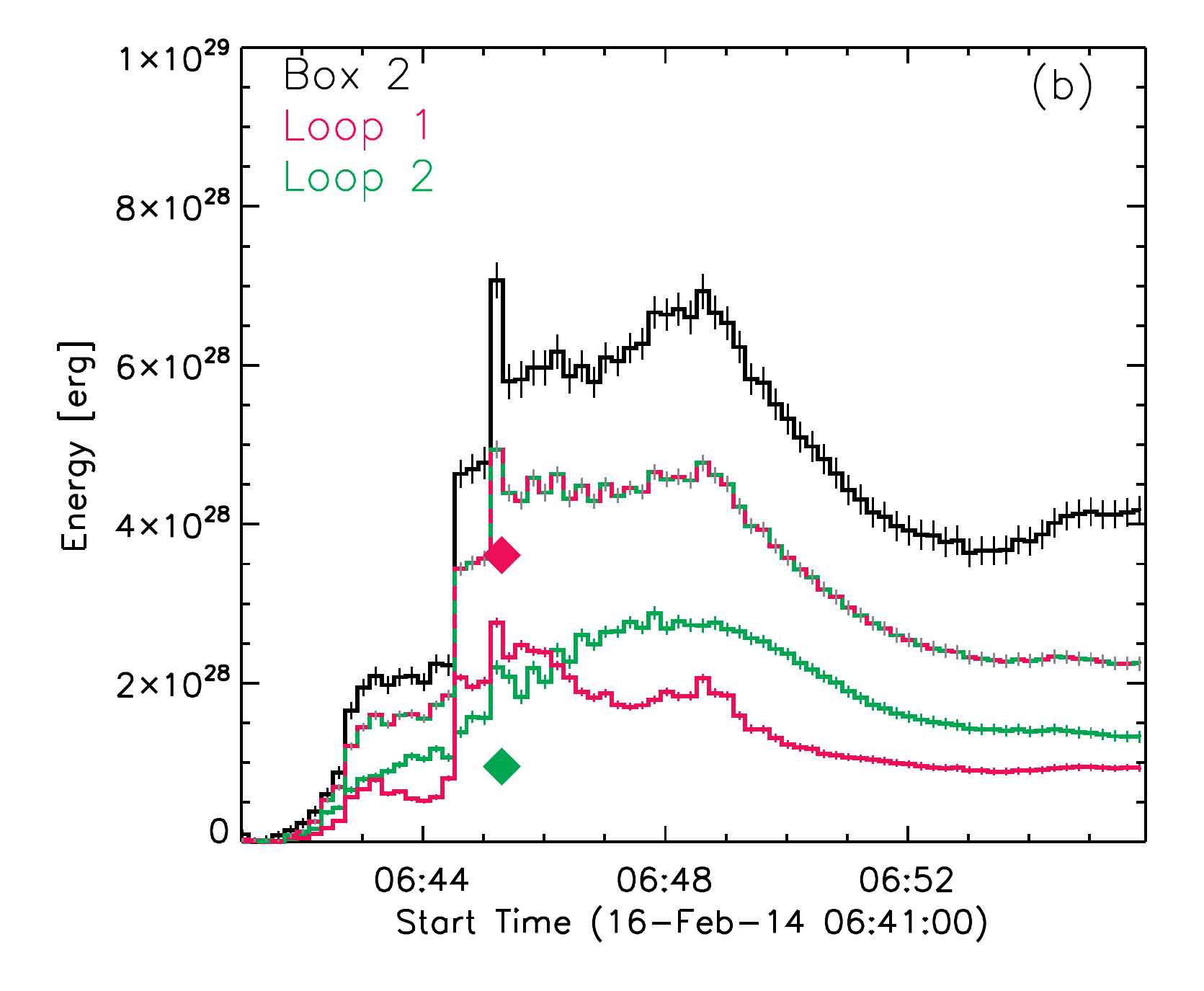}
\caption{
(a) Evolution of thermal energy computed from the DEM maps inside three boxes outlined by black, gray, and light gray contours in Figure\,\ref{Fig_EM_map}. (b) components of the thermal energy inside the black (flaring) box in Figure\,\ref{Fig_EM_map}: green/magenta lines correspond to green/magenta ROIs; the dashed green-magenta line shows the sum of them; the black line show the result for the entire box (same as the black line in panel a).
The red and green symbols indicate the model values of thermal energies Loops 1 and 2, respectively; see Table\,\ref{table_model_summary}.
\label{Fig_Energy_boxes}
}
\end{figure}

The total AIA-derived thermal energy of the flare in the FOV 
is obtained from the spatial distribution of the thermal energy density $w^{\rm{AIA}}_{ij}$,  computed from the regularized DEM maps in Section\,\ref{S_aia_diagn}, by adding up the contributions from all pixels in the FOV:
\begin{equation}\label{eq9}
W^{\rm{AIA}}_{\rm{therm}}(t)=S_{\rm{px}}\ l_{\rm{depth}}\sum_{i=1}^{N_{\rm{px}}} \sum_{j=1}^{N_{\rm{px}}} w^{\rm{AIA}}_{ij}(t) \; [\rm{erg}].
\end{equation}
{{The} animated Figure\,\ref{Fig_EM_map} shows that in addition to {the} evolution of the main flaring source, there is some dynamics in {the} top and bottom boxes (outlined in the Figure by gray and light gray contours). We cannot reliably conclude if this dynamics {is} related  to the flare or independent. To estimate contributions from those boxes to the thermal energy we show them along with the one from the middle (black) box in Figure\,\ref{Fig_Energy_boxes}a. The middle box uses the LOS depth $l_{\rm{depth}} =  3.8\times 10^8$\,cm, which is comparable to {the} width of the loops, while two other boxes employ $l_{\rm{depth}} =  7.6\times 10^7$\,cm, because the dynamic features are only about 1\arcsec\ wide. The Figure shows that the contributions from the top and bottom boxes are small compared with the main one; however, they display a behavior correlated in time with the main flaring box. This could indicate that the flaring process spreads out over a region larger than the main flaring loops. In what follows we, however, focus on the main flare region---the middle box.}

{Figure\,\ref{Fig_Energy_boxes}b displays further details of the thermal energy  distribution and evolution within the middle (black) box in Figure\,\ref{Fig_EM_map}. Here the green/magenta lines show the thermal energy computed for two different green/magenta ROIs in Figure\,\ref{Fig_EM_map}. These two ROIs are selected such as to inscribe the two different flaring loops as closely as possible. Note that this loop separation cannot be perfectly done because one of the loops projects onto the other.
The magenta and green symbols show the thermal energies computed by the volume integration of the thermal energy densities in flux tubes I and II, respectively. Although they agree with the data within a factor of two, they do not match each other perfectly. There are two possible causes of these mismatches: (i) the already mentioned projection effect and (ii) the ambiguity in determining the depth of the source, for which only a rough estimate is available from the data. Potentially, the values derived based on the 3D model are more precise as they are free from the LOS ambiguity.
The dashed green-magenta line shows the sum of those two contributions, while the black one shows the  thermal energy from the entire middle (black) box. This suggests that there is some thermal flare energy outside the most distinct loops, even though the loops give dominant contributions to the thermal energy budget.
}


{In all cases, a minimal preflare energy from the corresponding box or ROI was subtracted; for the middle (black) box} this minimal preflare value is $W^{\rm{AIA}}_{\rm{therm, min}}=2.54\times 10^{28}$\,[erg], which comes from the non-flaring pixels in the FOV. Evolution of this energy {(from the black box)} is shown in Figure\,\ref{Fig_Energy} in black. Its peak value is about $7\times 10^{28}$\,[erg].

\subsection{Thermal energy constrained with X-ray data and 3D modeling}

To calculate the thermal energy detected by \rhessi, we use the emission measure and temperature obtained from the \rhessi\ fit (to be specific, we only employ here the single-temperature + thick-target fit; see Section~\ref{S_rh_diagn} and green lines in Fig.~\ref{Fig_EM_T}; the alternative, two-temperature fit, yields very similar results):

\begin{equation}\label{eq6}
W^{\rm{RHESSI}}_{\rm{therm}}=3 k_B T_{\rm{RHESSI}} \sqrt{EM_{\rm{RHESSI}}\times V} \; [\rm{erg}],
\end{equation}
where $V$ is the volume of the corresponding thermal source. Here we use the model volume of Loop\,II from Table\,\ref{table_model_summary}. The evolution of this energy is shown in Figure\,\ref{Fig_Energy} in green; the peak value is about $5\times 10^{28}$\,[erg].

{We note that \rhessi\ is mostly sensitive to the hottest plasma within the FOV, namely for emission from flaring loop II in our case, which is also quantified by the AIA data in the previous section, see the green line in Figure\,\ref{Fig_Energy_boxes}b. It is instructive to compare these two green curves. The \rhessi-derived thermal energy has a peak around 06:47\,UT, where the temperature is the largest (about 30\,MK), while the AIA-derived thermal density of this loop has a peak one minute later, when the plasma cooled down to $\sim$20\,MK or less. The two energies agree well after that. The reason of the mismatch between these energies at the early decay phase is the well-known fact that AIA is not sensitive to plasma above $\sim$20\,MK; thus, AIA-based thermal diagnostics underestimates the thermal energy of that hot plasma, to which \rhessi\ is the most sensitive. }

\newpage

\subsection{Bulk kinetic energy {at the flare footpoints} }
\label{sec:energydensity}

Knowing the densities and the velocities, we can calculate the energy density $w_{bulk}^{\IRIS}$ as
\begin{equation}
\label{Eq_Doppl_Edens}
    w_{bulk}^{\IRIS} = 1.2 \cdot n_e  \cdot m_p \cdot \frac{v_{bulk}^2}{2} [erg\,cm^{-3}], 
\end{equation}
where $n_e$ is the electron density, the factor 1.2 appears because of ions, m$_p$ is the proton mass ($1.67\cdot10^{-24}$ g). We use the electron density determined from the \ion{O}{4} line ratios and the \ion{Si}{4} Doppler velocities because of the better Gaussian fits due to the higher signal and because \ion{Si}{4} velocities were found to be very similar to the \ion{O}{4} velocities.

We determined which \IRIS\ pixels lie inside the 50\% \rhessi\ 6-9 keV contour and calculated the energy density for all of them (where possible, unless the densities were unavailable). We then obtained an average energy density inside this area (0.12 erg cm$^{-3}$) {by averaging over the valid pixels}. 
To estimate the total bulk kinetic energy {within the flare footpoints} $W_{bulk}^{\IRIS}=\int w_{bulk}^{\IRIS} dV\sim w_{bulk}^{\IRIS} V$, we estimate the volume $V$ as a product of the number of pixels (1711) inside the contour and the pixel volume, assuming a height of 1 Mm. \IRIS\ observed with 0\farcs166/pixel = 120 km/pixel, giving a pixel area of $1.44 \cdot 10^{14}$ cm$^2$ and thus a volume of $1.44 \cdot 10^{22}$\,cm$^3$. This gives a total bulk kinetic energy  for this volume of  $W_{bulk}^{\IRIS}\sim3 \cdot 10^{24}$\,erg.

\subsection{Turbulent kinetic energy {at the flare footpoints}}

We determined the turbulent kinetic energy density $w_{turb}^{\IRIS}$ similarly to $w_{bulk}^{\IRIS}$ in Sect.~\ref{sec:energydensity} by replacing $v_{bulk}$ with $v_{turb}$ in Eq.~(\ref{Eq_Doppl_Edens}) and  considering the same valid pixels inside the 50\% \rhessi\ contour. The average turbulent energy density in this area is 2.7\,erg\,cm$^{-3}$. Similarly to Sect.~\ref{sec:energydensity}, we obtain the total energy by multiplying this energy density by the same volume as in Sect.~\ref{sec:energydensity}. 
The total turbulent kinetic energy  for this volume is then $W_{turb}^{\IRIS}\le 7 \cdot 10^{25}$ erg, which is less than 0.1\% of the flare thermal energy and about 0.15\% of the nonthermal energy deposition.
Neither $W_{bulk}^{\IRIS}$ nor $W_{turb}^{\IRIS}$ are shown in Figure\,\ref{Fig_Energy} as they are indistinguishable from the bottom axis.

We compare this turbulent energy detected in the flare footponts with the  turbulent energy  in a coronal flare volume.
\citet{2017PhRvL.118o5101K} employed the {\it Hinode}/EIS data to find the turbulent flare energy in a cusp region, which appeared to be less than 1\% of the released energy at any given time instance. The detected here kinetic turbulent energy at the flare footpoints, presumably driven by  precipitating particles, is one order of magnitude lower than that in the corona reported by \citet{2017PhRvL.118o5101K}.

\subsection{Quantification of the nonthermal energy with X-ray and \mw\ data and 3D modeling}

As previously mentioned, \rhessi\ missed the impulsive phase of the flare, while \kw\ recorded the impulsive phase in the G1 channel only in the waiting mode. Thus, nonthermal energy and the rate of the nonthermal energy deposition can only be estimated indirectly using the \mw\ spectrum and available \mw\ and \kw\ light curves. In Section\,\ref{S_nonthermal_population} we determined the population of nonthermal electrons needed to reproduce the observed peak \mw\ spectrum that occurred at 06:44:41\,UT. The corresponding instant nonthermal energy is $W_{nth}=5.87\times10^{27}$\,erg at that time.


Now, to estimate the nonthermal energy deposition rate we need to estimate the escape time $\tau_{\rm esc}$ of the nonthermal electrons from loop II. To do so, we consider lag-correlation between the HXR G1 \kw\ light curve and the \mw\ light curves at 3.75 and 9.4\,GHz obtained with NoRP. This lag-correlation analysis shows that the 9.4\,GHz light curve is delayed by $\sim$0.5\,s relative to the \kw\ light curve, while the 3.75\,GHz---by $\sim$2\,s. Thus, for our order-of-magnitude estimate, we select a characteristic value $\tau_{\rm esc}=1$\,s, which yields $\dot{W}_{nth}\approx5.87\times10^{27}$\,erg\,s$^{-1}$ at the flare peak time. To obtain the total deposition of the nonthermal energy, we multiply this peak value by the HXR light curve duration, 8\,s, at the level of half maximum: ${W}_{nth\,tot}\approx4.7\times10^{28}$\,erg. 

\begin{figure}\centering
\includegraphics[width=0.95\columnwidth]{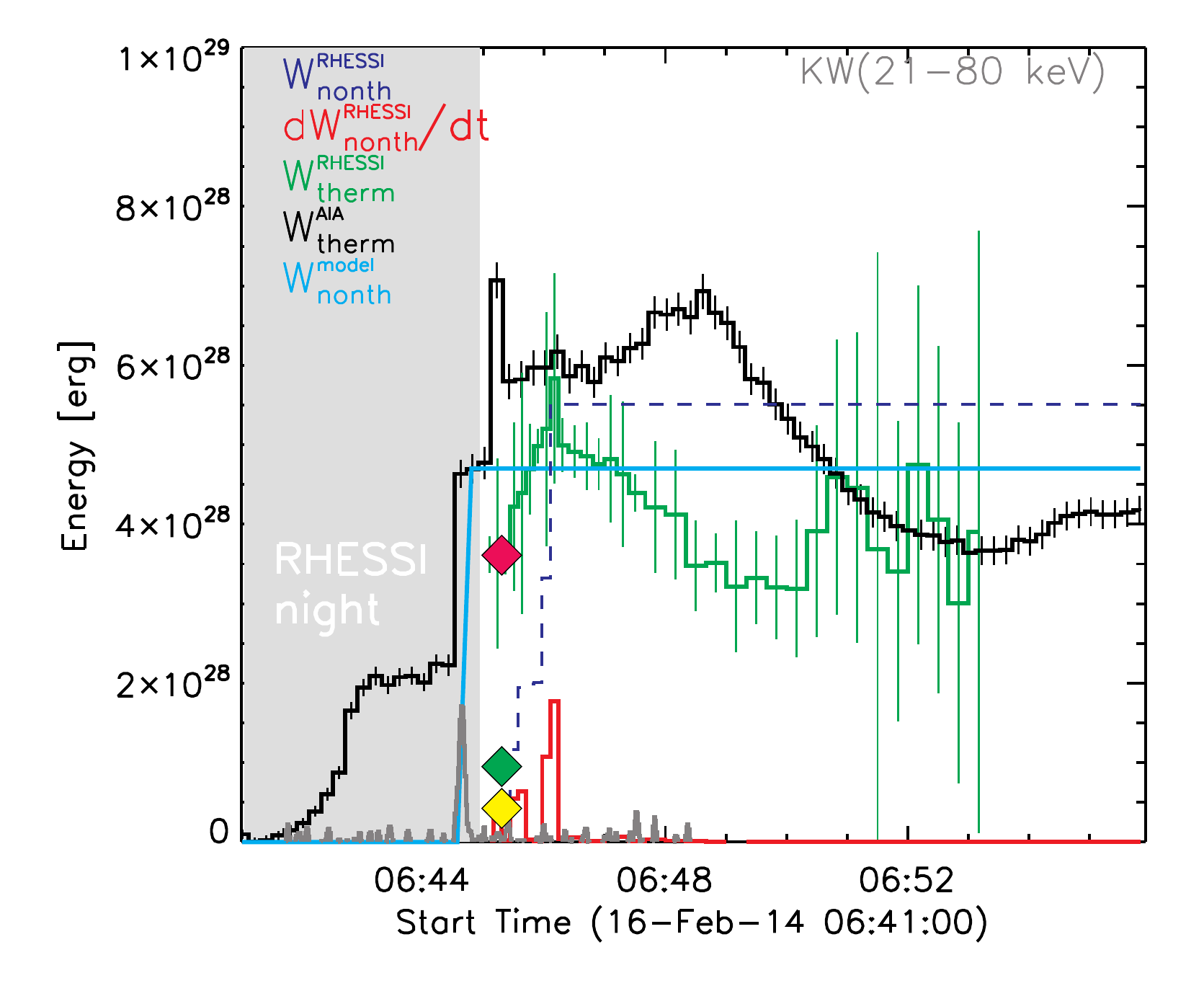}
\caption{ Evolution of energy components in the February 16, 2014 flare. Thermal energy $W^{\rm{RHESSI}}_{\rm{therm}}$ (Eq.\,\ref{eq6}) computed using thermal part of the \rhessi\ fits is shown in green, while the thermal energy  $W^{\rm{AIA}}_{\rm{therm}}$ (Eq.\,\ref{eq9}) computed from the {middle/black box of the} AIA DEM maps is shown in black. The grayed out portion of the plot indicates the \rhessi\ night. Normalized \kw\ light curve indicating the flare impulsive phase is shown in magenta, while the red histogram shows the rate of \rhessi\ nonthermal energy deposition $dW^{\rm{RHESSI}}_{\rm{nonth}}/dt$ [arb. units]. 
The nonthermal energy input inferred from the validated 3D model is shown in a light blue step-function.
The {red, green, and yellow} 
symbols indicate the model values of thermal energies in Loops 1, 2, and 3 respectively; see Table\,\ref{table_model_summary}.
Cumulative nonthermal energy deposition  $W^{\rm{RHESSI}}_{\rm{nonth}}$ after the impulsive phase  obtained using parameters of the nonthermal part of the \rhessi\ fits (Eq.\,\ref{eq5}) is shown in dashed blue histogram.
\label{Fig_Energy}
}
\end{figure}

The nonthermal energy $W^{\rm{RHESSI}}_{\rm{nonth}}$ potentially captured by \rhessi\ after the impulsive phase was computed as a cumulative sum using the parameters from the \rhessi\ fits: 
\begin{equation}\label{eq5}
W^{\rm{RHESSI}}_{\rm{nonth}}=\int\limits_{t_0}^t F_0 E_c \frac{\delta-1}{\delta-2} dt \; [\rm{erg}],
\end{equation}
where $t_0$ is the end of the \rhessi\ night, while the  $F_0$, $E_c$, and $\delta$ are the thick target parameters (see Section\,\ref{S_rh_diagn}) from the fitting of the \rhessi\ data displayed in Figure\,\ref{Fig_EM_T}c-e;  $W^{\rm{RHESSI}}_{\rm{nonth}}$ is shown in dashed blue in Figure \ref{Fig_Energy}. We note that these thick-target fit parameters, especially $F_0$, come with very large uncertainties and, might overestimate the nonthermal energy deposition. However, as has been shown, the presence of such a nonthermal population does not contradict any available data. For example, the \mw\ emission produced by such a population does not overestimate the observed \mw\ emission. The presence of this nonthermal component is also consistent with the behaviour of the thermal energy at the hot loop, which reaches the peak exactly when this nonthermal component is over. 


\subsection{Evolution of the energy partitions}

{Here we briefly summarize  the evolution of the energy partitions at the course of the three phases of the flare---the pre-impulsive, the impulsive, and the main phase. We only consider Loops I and II as we do not have enough constraints to quantify evolution of Loop III. At the pre-impulsive phase, we observe a modest heating of Loops I ($T\sim9-10$\,MK) and II ($T\sim13-14$\,MK). 
No signature of any nonthermal component is present at this phase. This phase is followed by a very short (20\,s) impulsive phase, when the release of the nonthermal energy observed in the \mw\ and HXR ranges are immediately followed by a thermal response in Loops I and II. The heating continues for a few more minutes into the main phase of the flare. Both pre-impulsive and main flare phases require an extra heating in addition to the heating due to nonthermal electron loss. }

\newpage
\subsection{Magnetic energy available for the flare}

The described above components of the flare energy must have come from the free magnetic energy available in the AR \citep{2020Sci...367..278F}. To check if the needed magnetic energy existed in the AR before the flare and nail down its changes associated with the flare we resort to 3D magnetic reconstructions. To do so we fixed the Carrington coordinates of our model FOV and created sequences of potential and NLFFF reconstructions computed for seven consecutive time frames between 6:00 and 7:12\,UT roughly centered at the flare time. This approach preserves the reconstruction volume for different time frames and minimizes possible uncertainties due to solar rotation. We employed two different disambiguation algorithms for the transverse component of the photospheric magnetic field \citep{1994SoPh..155..235M,2014SoPh..289.1499R} and found only very minor differences between solutions, which cross-validates them.

The total magnetic energy in each data cube is straightforwardly  computed as $W_{B}=\int B^2 dV/(8\pi)$, which fluctuates less than 1\% around  $\approx1.2\times10^{33}$\,erg for all our potential extrapolations and around $\approx1.1\times10^{33}$\,erg for the NLFFF extrapolations.
Therefore, we cannot derive the free magnetic energy because the magnetic energy in all NLFFF cubes appeared slightly smaller, by less than 10\% in all cases, than the corresponding potential magnetic energy.  This is a known problem of reconstruction algorithms associated with numerical residuals of $\nabla\cdot \mathbf{B}$, which are theoretically equivalent to zero, but in the numerical solutions---not \citep{2015ApJ...811..107D}.

We estimated uncertainties of the total magnetic energy as follows: (i) by finding differences of the total magnetic energy between two successive (in time) data cubes and (ii) by finding differences between two simultaneous models extrapolated from the bottom boundary condition obtained with two different  $\pi$-disambiguation methods \citep{1994SoPh..155..235M,2014SoPh..289.1499R}. The magnetic energy uncertainties estimated this way are above $\sim3\times10^{30}$\,erg. This  is more than one order of magnitude above the required flare energy, $\sim10^{29}$\,erg; thus, the decrease of the magnetic energy due to the flare cannot be detected using the NLFFF extrapolation method.
The total energy released in the flare comprises only less than  0.01\% of the total magnetic energy.
We conclude that the presence of the required free magnetic energy  in this AR is highly likely.

\section{Discussion}

In this study, we have analysed energy partitioning, its evolution,
and spatial distribution in flaring loops. We found three distinct consecutive heating episodes, while only one single episode of nonthermal emission. The first {pre-}heating episode (a precursor?) preceded the nonthermal impulsive peak, the second one appeared as a response on the impulsive peak, while the third one occurred later {at the main flare phase} without any obvious connection with the {impulsive} nonthermal peak. From the timing of the event, we can conclude that only a portion of the flaring plasma heating was driven by nonthermal electron losses, while the remaining portion was driven by another agent, which could be, e.g., a direct heating by magnetic reconnection or heating by accelerated ions or something else.

The distribution of the energy components over the three flaring loops involved in the event is highly uneven. While the plasma heating takes place in all three loops, the nonthermal electrons are present in one loop only---that showing the strongest plasma heating. The nonthermal energy deposition is sufficient for the post-impulsive heating in  the flaring loop with nonthermal electrons, while insufficient for the first and third heating episodes.

The flare-integrated energies are: (i) the thermal energy $\sim10^{29}$\,erg;  (ii) the nonthermal energy deposition $\sim5\times10^{28}$\,erg; (iii) the turbulent kinetic energy at the footpoints $\sim7\times10^{25}$\,erg; (iv) the bulk kinetic energy at the footpoints $\sim3\times10^{24}$\,erg. These comprises less than only 0.01\% of the total magnetic energy of the AR.

{The kinetic energy detected in our study, presumably driven by nonthermal electrons precipitating at the footpoins, represents only a minor fraction of the released energy and so energetically unimportant. We note that  kinetic energy of plasma motions could be measured in UV, EUV, and soft-X-rays
and the different spectral lines tell us about moving plasma at different temperatures and
in different layers of flaring atmosphere. So the kinetic energy (energy density) estimates
could be different for different parts of a flare. With \IRIS\ UV data we have only estimated a portion of the kinetic energies in the flare footpoint, while this energy remains unconstrained in the flare coronal volume. The only available coronal spectral line in our case is \ion{Fe}{21}, sensitive to the 10\,MK plasma, which is weak and not suitable for quantitative fitting with a gaussian spectral profile. In principle, some portion of the kinetic energy in the corona could be estimated from apparent motions of the hot flaring plasma, which are, however, not seen in the animated Figure\,\ref{Fig_EM_map}, where all apparent changes indicate a temporal evolution of the plasma temperature and emission measure, rather than any change of the source morphology, which could be attributed to plasma motions. This means that no available diagnostic reveals any measurable coronal kinetic energy in this flare.  }

The flare properties summarised above place our event somewhere in between ``normal'' flares with a thermal precursor and nonthermally-dominated (``cold'') flares. The estimated total deposition of the nonthermal energy is only a factor of two lower than the detected thermal energy; considering uncertainties of these estimates this might be sufficient to account for the thermal energy of the flare in agreement with a conclusion of  \citet{2012ApJ...759...71E} obtained for a set of powerful flares. However, the detailed analysis of timing of thermal and nonthermal emissions as well as of spatial placements of the thermal and nonthermal energies between the three flaring magnetic flux tubes rules out such an option. Meanwhile, one of the flaring loops, Loop II in our model, shows close similarity to the nonthermally-dominated ``cold'' flares: there is a clear Neupert effect timing between the impulsive nonthermal emission and thermal response in Loop II, although even here a preheating is detected. The nonthermal energy deposition matches well the thermal energy detected in this flare, while the associated kinetic energy from the flare footpoints adds only a minor fraction to the overall energy budget. This implies that the nonthermal energy is primarily dissipated to the thermal energy rather than to bulk or random motions of the ambient plasma.

For comparison, in the nonthermally-dominated 2013-Nov-05 flare described in Paper I, all (two) flaring loops contain accelerated nonthermal electrons and are heated due only to losses of those nonthermal electrons.
The nonthermal electrons were divided roughly equally between those loops, but the thermal responses of those loops were different from each other. \citet{2020ApJ...890...75M} proposed that the thermal response of a loop differs depending of the initial plasma density in the loop just before the episode of the energy release.
In the 2014-Feb-16 flare, studied here, we observe more dissimilarity between the loops involved in the flaring. Only one of them contains a detectable amount nonthermal electrons (at the peak time) and only in that loop where the heating is dominated by losses of the nonthermal electrons. Two other flaring loops are heated, presumably, by some other means. We note that the initial densities in the loops are different from each other. The nonthermal electrons are observed in the most tenuous and hottest of the three loops. This loop is heated up to highest temperatures compared with the ``directly'' heated loops. A similar behaviour, with the largest temperature detected in the most tenuous loop, is detected in the nonthermally-dominated 2013-11-05 flare. We, therefore, propose that not only the magnetic structure in the flaring volume, but also initial distribution of the thermal plasma plays a crucial role in deciding how the released energy is apportioned between the thermal and nonthermal components and how these energies are divided between various flaring loops.

\acknowledgements
This work was partly supported 
by NSF AGS-1817277 
and AGS-1743321 
grants
and NASA grants
80NSSC18K0667, 
80NSSC19K0068, 
80NSSC18K1128, 
and 80NSSC20K0627 
to New Jersey Institute of Technology.
GM (\sdo/AIA and \rhessi\; data analysis) acknowledges support from RSF grant 20-72-10158.
EPK was supported by STFC consolidated grant  ST/P000533/1. {We are thankful to Alexandra Lysenko for fruitful discussions of the \kw\ data.}

\newpage

\bibliography{all_issi_references,2017sep10,fleishman}

\end{document}